\documentclass[journal]{IEEEtran}
\ifCLASSINFOpdf
\else
\fi
\usepackage{cite}
\usepackage{amsmath}
\usepackage{amssymb}
\usepackage{graphicx}
\usepackage{epstopdf}
\usepackage{setspace}
\usepackage{stfloats}
\usepackage{makecell}
\usepackage{amsfonts,amssymb}
\usepackage{mathrsfs}
\usepackage{algorithmic}
\usepackage{algorithm}
\usepackage{float}
\usepackage{color}
\usepackage{enumerate}
\newtheorem{remark}{Remark}
\newtheorem{lemma}{Lemma}
\newtheorem{proof}{Proof}[section]
\begin{document}
\title{Link Selection for Secure Cooperative Networks with Buffer-Aided Relaying}
\author{Ji He,~\IEEEmembership{Student~Member,~IEEE,}
        Jia Liu,~\IEEEmembership{Member,~IEEE,}
				Yulong Shen,~\IEEEmembership{Member,~IEEE}
        and~Xiaohong~Jiang,~\IEEEmembership{Senior~Member,~IEEE}
}

\maketitle

\begin{abstract}
This paper investigates the secure communication in a two-hop cooperative wireless network, where a buffer-aided relay is utilized to forward data from the source to destination, and a passive eavesdropper attempts to intercept data transmission from both the source and relay. Depending on the availability of instantaneous channel state information of the source, two cases of transmission mechanisms, i.e., adaptive-rate transmission and fixed-rate transmission are considered. To enhance the security of the system, novel link selection policies are proposed for both cases to select source-to-relay, relay-to destination, or no link transmission based on the channels qualities. Closed-form expressions are derived for the end-to-end secrecy outage probability (SOP), secrecy outage capacity (SOC), and exact secrecy throughput (EST), respectively. Furthermore, we prove the condition that EST reaches its maximum, and explore how to minimize the SOP and maximize the SOC by optimizing the link selection parameters. Finally, simulations are conducted to demonstrate the validity of our theoretical performance evaluation, and extensive numerical results are provided to illustrate the efficiency of the proposed link selection polices for the secure communication in two-hop cooperative networks.
\end{abstract}

\IEEEpeerreviewmaketitle

\section{Introduction}

  \IEEEPARstart{W}{ireless} communication technologies are the fastest growing segment of the telecommunications industry and their rapid developments have been promoting the evolution into the fifth generation (5G) communication \cite{16COMST:5G}. However, due to the broadcast nature of wireless mediums, communication over wireless networks is susceptible to the interception attacks of unintended recipients (i.e., eavesdroppers). Therefore, guaranteeing the security of wireless communication networks is becoming an increasingly urgent demand \cite{02NIST:wireless_network_security}.

  Traditionally, data is secured by applying the key-based enciphering (cryptographic) techniques in the upper layers of the network protocol stack \cite{14book:cryptography}. Although these cryptographic methods have shown their effectiveness in wired networks, the inherent difficulty of secret key distribution/management without centralized control and complex encryption algorithms may significantly limit their applications in decentralized wireless networks \cite{1998Computer:CrypVulnerabilities}. This motivates the introduction of physical layer (PHY) security technology recently as the complementary approach to further enhancing the secrecy in wireless communications \cite{11book:PHYsecurity}.  The philosophy behind PHY security is to exploit the natural randomness of noise and the physical characteristics of wireless channels (like fading) to provide information-theoretic security, which has been regarded as the strongest form of security irrespective of the computing capabilities of eavesdroppers \cite{06IT:SCofwirelesschannels,08TIT:WirelessInformation-TheoreticSecurity,14COMST:PrinciplesofPHYSecurity}. Thus, PHY security techniques are highly promising to guarantee everlasting secure communication for wireless networks \cite{08TIT:fadingchannel,15CM:PHY-multi-antennas,15CM:PLSin5G}.

  The seminal work \cite{1975:wire-tapchannel} by Wyner introduced the wiretap channel model as a basic framework for the PHY security which was laid down by Shannon's definition of perfect secrecy in \cite{1949:shannoncommunication}. Subsequently, many research activities have been devoted to the study of PHY security under other channel models, such as non-degrade channel \cite{1978TIT:broa-channel}, Gaussian channel \cite{1978TIT:Gau-channel}, multi-antenna channel \cite{05TIT:multi-channel} and relay channel \cite{08TIT:relay-channel}. Motivated by these early studies, diverse approaches for improving PHY security have been proposed in the literature, which mainly include channel precoding/beamforming \cite{10TIT:MISOME-bea,10TIT:MIMOME-bea,17TSP:precoding}, cooperative jamming \cite{08TIT:two-wayjamming,11TSP:MIMO-jamming}, channel coding \cite{14WCL:coding1,11TIFS:LDPCcoding} and link/relay selection \cite{06JSAC:linkselection,12TOC:relayselection,15TWC:linkselection}.

  This paper focuses on the link/relay selection for securing the communication in wireless cooperative networks. The main advantage of link/relay selection is the implementation simplicity, as the sophisticated transmission techniques or explicit synchronization process is not required. Inspired by the pioneering work \cite{06JSAC:linkselection} which analyzes the fundamental benefits can be achieved by link/relay selection, extensive studies have been conducted to design efficient relay/link selection schemes under various network scenarios, such as max-min-ratio scheme \cite{12TOC:relayselection}, AFbORS and DFbORS schemes \cite{15TWC:linkselection}. Recently, the adoption of buffer-aided relaying has been proved that it can improve the cooperative communication performance in terms of throughput and diversity gains \cite{13TIT:buffer-aided,15CL:buffer-aided}. Different from the conventional relaying scheduling, buffer-aided relaying exploits the flexibility offered by the buffer and enables the data transmission to be executed under favorable channel conditions.

  Following this line, some initial selection schemes with buffer-aided relaying have been proposed for the secure communication in wireless cooperative networks \cite{14TIFS:max-ratio,15TWC:linkselectionbuffer,15WSA:buffer,AN_aided,16CL:linkselection}. Specifically, Chen \emph{et al.} \cite{14TIFS:max-ratio} put forward the max-ratio (MR) selection scheme for half-duplex decode-and-forward (DF) relaying networks. MR scheme activates the link with the largest channel gain ratio based on the knowledge of both legitimate and wiretap channel state information (CSI), such that it can achieve a better secrecy performance than the conventional max-min-ratio scheme \cite{12TOC:relayselection}. Taking into account the transmission efficiency and security constraint, Huang \emph{et al.} \cite{15TWC:linkselectionbuffer} designed a link selection scheme in a two-hop DF relay network to achieve tradeoff between secrecy throughput and secrecy outage probability. The network scenario with multiple antennas and multiple eavesdroppers was further explored in \cite{15WSA:buffer}, where a maximum likelihood (ML) criterion-based algorithm is proposed to select sets of relays for secure transmission. More recently, an artificial noise injection scheme and a hybrid half-/full-duplex scheme were investigated in \cite{AN_aided} and \cite{16CL:linkselection}, respectively, to enhance the physical layer security in cooperative networks with buffer-aided relaying.

  This paper considers a more practical wireless cooperative system which composes one source-destination pair, one trusted relay with infinite buffer and one passive eavesdropper. Taking into account the fact that the eavesdropper intercepts data transmission in a passive way which can be hardly monitored, we adopt the assumption that the exact instantaneous/statistical CSI of the eavesdropping channel is \emph{unavailable}, reflecting the more realistic scenario than those assumed in the aforementioned works. Moreover, in contrast to \cite{15TWC:linkselectionbuffer} where only the relay-to-destination channel is wiretapped, we consider a more hazardous and more practical eavesdropping scenario that the eavesdropper overhears data transmission from both the source-to-relay and relay-to-destination links in order to intercept the confidential information with a higher probability. Therefore, to ensure the end-to-end secure communication in wireless cooperative networks, novel link selection schemes should be redesigned, which motivates the conducting of this study. The main contributions of this paper are summarized as follows:
   \begin{itemize}
     \item Depending on the availability of instantaneous channel state information at the source, two cases of transmission mechanisms, i.e., adaptive-rate transmission and fixed-rate transmission are considered. We design link selection policies to ensure the communication security for both cases. Particularly, according to the qualities of legitimate channels, the policies fully utilize the flexibility provided by buffer-aided relaying to select source-to-relay, relay-to-destination, or \emph{no link transmission}, which are different from the conventional on-off schemes.

     \item For the proposed link selection policies, closed-form expressions of end-to-end secrecy outage probability (SOP), secrecy outage capacity (SOC) and exact secrecy throughput (EST) are derived, respectively. We further prove the condition that EST reaches its maximum, and explore how to minimize the SOP and maximize the SOC by optimizing the link selection parameters.

    \item We conduct simulations to demonstrate the validity of theoretical performance evaluation, and provide extensive numerical results to illustrate the efficiency of the proposed link selection polices for the secure communication in wireless cooperative networks.
	\end{itemize}
	
  The remainder of this paper is outlined as follows. Section~\ref{section:system_model} introduces the system models and necessary definitions. Section~\ref{sec:link selection scheme} elaborates the design of link selection policies under the adaptive-rate and fixed-rate transmission mechanisms, respectively. The general problem formulations are presented in Section~\ref{section:problem_formulation}£» Section~\ref{section:performance_evaluation} conducts the performance evaluation and explores the performance optimization issues£» We provide simulation and numerical results in Section~\ref{section:simulation} and finally concludes this paper in Section~\ref{section:conclusion}.

\section{System Models and Definitions} \label{section:system_model}
  In this section, we introduce the system models and some definitions involved in this study.

 \subsection{Network Model} \label{subsec:network model}
\begin{figure}[t]
\centering
\includegraphics[width=1\linewidth]{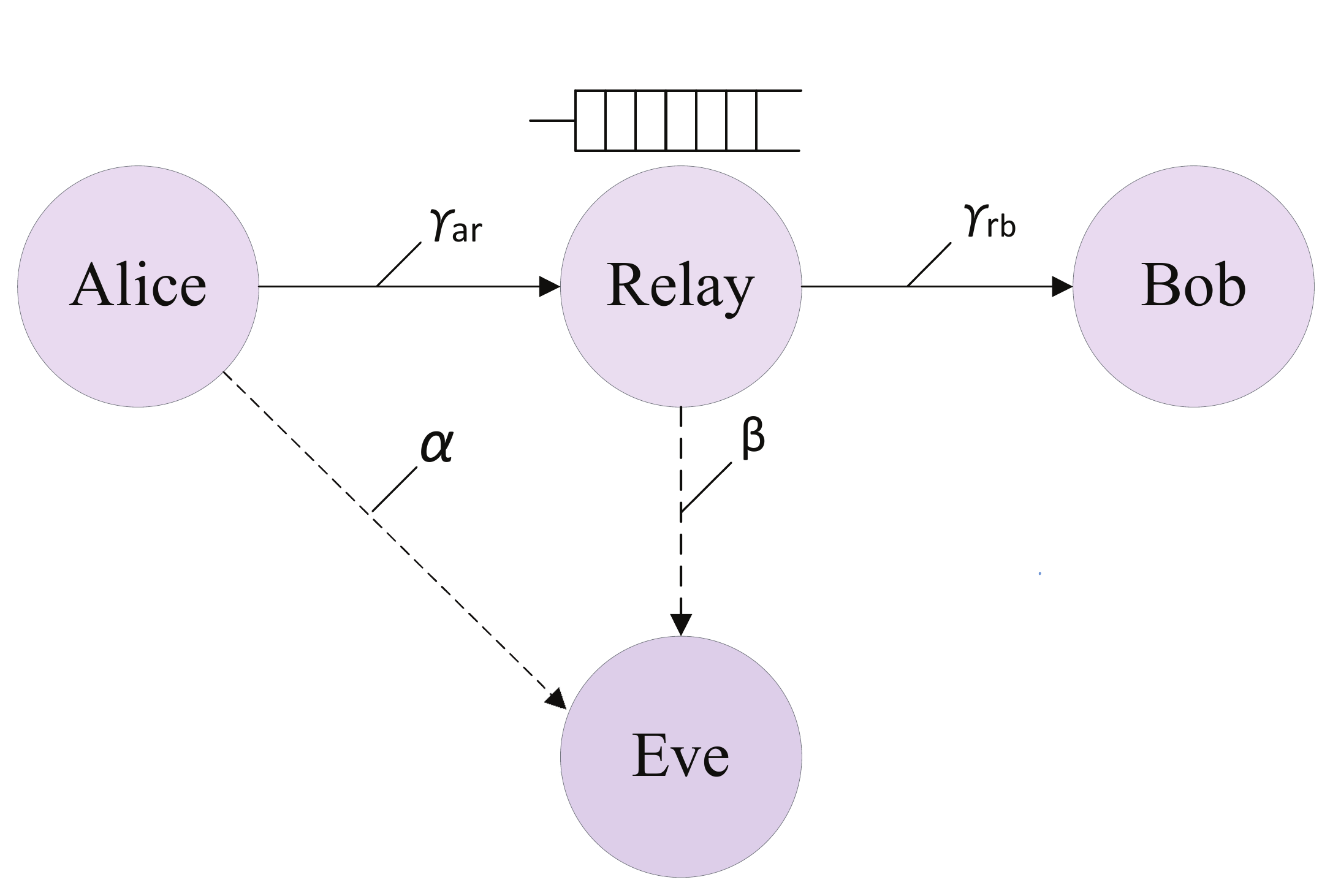}
\caption{Illustration of system models.}
\label{sysmodel}
\end{figure}

  As shown in Fig.~\ref{sysmodel}, we consider a two-hop wireless cooperative network which consists of a source (Alice), a destination (Bob), a relay (Relay) and a passive eavesdropper (Eve). We assume that there is no direct link from Alice to Bob so that the messages from Alice can be delivered to Bob only via Relay. Relay is equipped with infinite buffer to temporarily store the messages from Alice and operates in the half-duplex mode, thus it can not transmit and receive simultaneously. Moreover, we apply the randomize-and-forward (RF) strategy \cite{12CL:RF}, with which Relay decodes the original signal from Alice and store the message in its buffer, later it forwards the message to Bob by transmitting independent randomization signal. We assume that Alice and Relay transmit messages with fixed power $P_a$ and $P_r$, respectively. Eve attempts to intercept signals from both Alice and Relay, but  it cannot combine signals of the two hops with combining techniques such as MRC \cite{04TIT:MRC,14TWC:MRC} due to the RF strategy.

 \subsection{Wireless Channel Model} \label{subsec:channel model}
  We consider a time-slotted system where the time is divided into successive slots with equal duration. All wireless links are characterized by the quasi-static Rayleigh block fading such that the channel fading coefficient of each link remains constant during one time slot, but changes independently and randomly from one time slot to the next. We use $h_{i,j}[k]$ to denote the fading coefficient from node $i$ to node $j$ at time slot $k$, where $i \in \{a,r\}$ and $j \in \{r,b,e\}$ (here $a$, $r$, $b$, $e$ are short for Alice, Relay, Bob and Eve, respectively). With the quasi-static Rayleigh block fading model, the channel gain of a link is independently and exponentially distributed with mean $\mathbb{E} \{|{h_{i,j}[k]}|^2\}={\Omega _{i,j}}$, where $\mathbb{E}\{\cdot\}$ is the expectation operator. In addition, complex additive white Gaussian noise (AWGN) is imposed on each link and its variance at Relay, Bob and Eve are $\delta _{r}^2$, $\delta _{b}^2$ and $\delta _{e}^2$, respectively. Therefore, the instantaneous signal-to-noise ratio (SNR) $\gamma _{i,j}[k]$ of a link at time slot $k$ is determined as
\begin{equation}
  \gamma _{i,j}[k] = \frac{{P_i}}{\delta _j^2}|h_{i,j}[k]|^2. \label{eq:snr}
\end{equation}
  $\gamma _{i,j}[k]$ is also exponentially distributed with the probability density function (p.d.f) given by
\begin{equation}
  f_{\gamma _{i,j}[k]}(x) = \frac{1}{\bar{\gamma}_{i,j}} \text{exp}\left(-\frac{x}{\bar{\gamma}_{i,j}}\right), x \ge 0,
  \label{eq:pdf}
\end{equation}
  where $ \bar \gamma _{i,j} = \frac{P_i}{\delta _j^2}\Omega _{i,j}$. Considering the fact that Eve is a passive eavesdropper, the instantaneous CSIs from Alice and Relay to Eve, i.e., $h_{a,e}[k]$ and $h_{r,e}[k]$, are unavailable. Moreover, in this study we assume that Relay always knows the instantaneous CSI at Bob while Alice may or may not know the instantaneous CSI at Relay, as explained later.

\subsection{Transmission Mechanism} \label{subsec:transmission mechanism}
  To guarantee the secrecy transmission, we employ the well-known Wyner's encoding scheme \cite{1975:wire-tapchannel}. When a transmission is conducted, the transmitter (Alice or Relay) chooses two rates, one is the rate of transmitted codewords, another is the rate of confidential messages. The difference between the two rates, i.e., the rate redundancy, reflects the cost of the secrecy transmission against eavesdropping. Since we consider the practical scenario that the instantaneous/statistical CSI of the wiretap channel is unknown, the transmitter cannot determine the cost needed to prevent eavesdropping. As a result, in our transmission mechanism, Alice and Relay set a fix rate for the confidential messages, denoted as $R_s$.

  Regarding the transmission from Alice to Relay, we consider two cases, i.e., Alice knows and does not know the corresponding instantaneous CSI (the availability of instantaneous CSI at Alice is dependent on the link selection policy adopted, as explained in Section~\ref{sec:link selection scheme}). For the former case, Alice adaptively adjusts the codeword rate to be arbitrarily close to the channel capacity \cite{13TIT:channeloutage}, termed as \emph{adaptive-rate transmission}. Thus, when Alice-to-Relay link is selected in time slot $k$, the transmission rate of codewords $R_{a,r}[k]$ is determined as
\begin{eqnarray}
 {R_{a,r}}[k] = \log_2(1 + \gamma _{a,r}[k]).
 \label{eq:Rar}
\end{eqnarray}
  For the case that Alice does not know the instantaneous CSI, when Alice-to-Relay link is selected, it sets a fix rate $R_a$ ($R_a \geq R_s$) to transmit the codewords, termed as \emph{fixed-rate transmission}.

  Regarding the transmission from Relay to Bob, Relay always knows the corresponding instantaneous CSI based on the link selection policies as introduced later. Thus, when Relay-to-Bob link is selected in time slot $k$, the transmission rate of codewords $R_{r,b}[k]$ is determined as
\begin{eqnarray}
 {R_{r,b}}[k] = \log_2(1 + \gamma_{r,b}[k]).
 \label{eq:Rrb}
 \end{eqnarray}

  We use $Q_r[k-1]$ to denote the amount of confidential data (in bits) stored in the buffer of Relay at the end of time slot $k-1$. Then, the evolution of data stored in Relay's buffer at the next time slot can be characterized as\footnote{It should be noted that after decoding the signal from Alice, Relay only stores the useful data, i.e., the confidential messages, in its buffer.}
\begin{equation}
Q_r[k]=\begin{cases}
Q_r[k-1]+R_s,        &\text{Alice-to-Relay is selected}\\
\{Q_r[k-1]-R_s\}^+,  &\text{Relay-to-Bob is selected}\\
Q_r[k-1],            &\text{No link is selected}
\end{cases}
\label{eq:Q_r_k}
\end{equation}
where $\{x\}^+=\max\{x,0\}$.

\begin{remark}
  As introduced in the next section, our link selection polices can guarantee that the rate of codewords is always no less than the rate of confidential messages $R_s$. More specifically, when Alice-to-Relay link is selected in time slot $k$ and the adaptive-rate transmission is adopted, $R_{a,r}[k] \geq R_s$ always satisfies; when Relay-to-Bob link is selected in time slot $k$, $R_{r,b}[k] \geq R_s$ always satisfies.
\end{remark}



\section{Link Selection Policies} \label{sec:link selection scheme}
In this section, we first present the overall scheduling in a time slot, and then detail the link selection policies.

\subsection{Overall Scheduling}

\begin{figure}[t]
\centering
\includegraphics[width=1\linewidth]{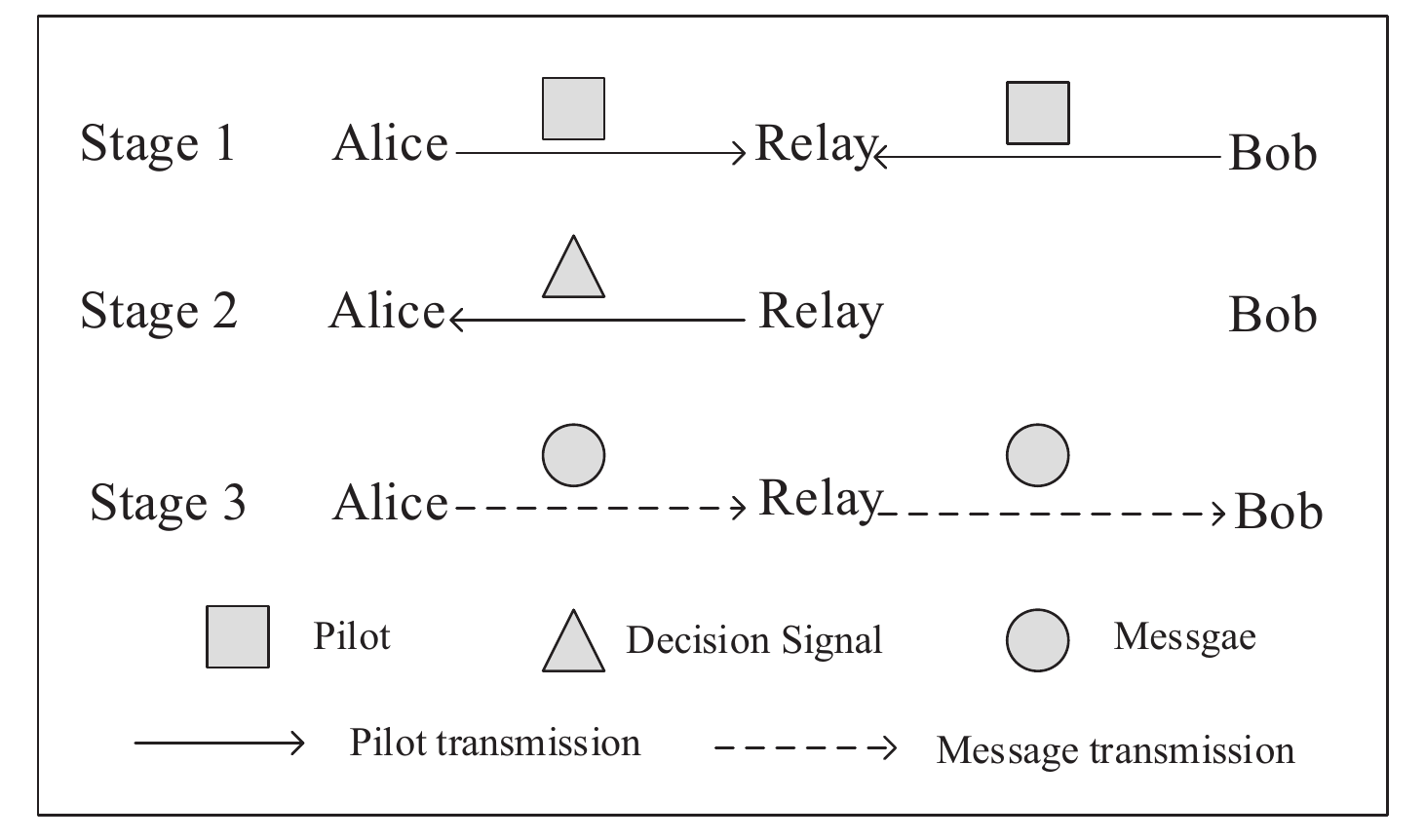}
\caption{Illustration of overall scheduling in a time slot.}
\label{fig:overall_scheduling}
\end{figure}

Regarding the overall scheduling in a time slot, in order to ensure the transmission security and avoid channel outage \cite{13TIT:channeloutage}, we first need to estimate the instantaneous CSIs of legitimate links. Then, link selection is made based on our new policies. Finally, the system conducts transmission operation or \emph{remains idle} according to the selection decision. Therefore, the overall scheduling consists of the following three stages and can be illustrated in Fig.~\ref{fig:overall_scheduling}.

\begin{enumerate}[{Stage} 1]

\item
(\textbf{CSI Estimation})

Alice and Bob transmit the pilot sequences to Relay in turn. By assuming that the reciprocity property \cite{1943:reciprocity} of antenna holds, Relay can estimate the CSIs of Alice-to-Relay and Relay-to-Bob links, respectively.

\item
(\textbf{Link Selection})

With the CSIs of two links, Relay acts as the `central node' to make link selection decision based on some policies. According to whether Relay feed back the CSI to Alice, we consider the following two cases.

\begin{itemize}

\item[\textbf{a)}]
\emph{With CSI feedback}: Relay makes link selection decision based on the policy described in Section~\ref{section:policy_with_CSI}. If Alice-to-Relay link is selected, Relay sends the decision signal and feeds back the CSI to Alice.

\item[\textbf{b)}]
\emph{Without CSI feedback}: Relay makes link selection decision based on the policy described in Section~\ref{section:policy_without_CSI}. If Alice-to-Relay link is selected, Relay sends the decision signal to Alice.

\end{itemize}

\item
(\textbf{Message Transmission})

Based on the link selection decision, Alice or Relay transmits the message with the transmission mechanism introduced in Section~\ref{subsec:transmission mechanism}, or the system remains idle.

\end{enumerate}

\begin{remark}
It is worth noting that the overall scheduling incurs at most three handshakes before the real message transmission, thus it is low-complexity for the system operation.
\end{remark}

\subsection{Link Selection Policy with CSI Feedback} \label{section:policy_with_CSI}

With the existing link selection policies such as \cite{15TWC:linkselectionbuffer}, either Alice-to-Relay or Relay-to-Bob link is selected for data transmission in all time slots. However, since the eavesdropper Eve intercepts messages from both links, once in a time slot the channel qualities of both legitimate links are worse than those of corresponding wiretap links, the transmission security cannot be ensured no matter which link is selected. Therefore, a new selection policy with such a consideration should be carefully designed.

We let $I_k$ be an indicator variable to denote the link decision in time slot $k$. $I_k=0$ indicates Alice-to-Relay link is selected, $I_k=1$ indicates Relay-to-Bob link is selected, and $I_{k}=-1$ indicates the system remains idle in this time slot. We also introduce two non-negative parameters $\alpha$ and $\beta$ which serve as the thresholds for the channel qualities of two legitimate links, respectively. Specifically, only if the condition $\gamma_{a,r}[k] \geq \alpha$ (resp. $\gamma_{r,b}[k] \geq \beta$) satisfies, Alice-to-Relay (resp. Relay-to-Bob) link can be selected for message transmission, if $\gamma_{a,r}[k] < \alpha$ and $\gamma_{r,b}[k] < \beta$, no link will be selected, which ensures a high channel quality of the selected link and thus provides a good security performance. Moreover, in order to guarantee that the rate of codewords of the selected link can cover the rate of confidential messages $R_s$, we set $\alpha \geq 2^{R_s}-1$ and $\beta \geq 2^{R_s}-1$. Finally, when both the legitimate links are in high channel quality, i.e., $\gamma_{a,r}[k] \geq \alpha$ and $\gamma_{r,b}[k] \geq \beta$, the link with a better relative quality will be selected for message transmission, i.e., $I_k=0$ if $\frac{\gamma_{a,r}[k]}{\alpha} \geq \frac{\gamma_{r,b}[k]}{\beta}$ and $I_k=1$ if $\frac{\gamma_{a,r}[k]}{\alpha} < \frac{\gamma_{r,b}[k]}{\beta}$. Therefore, our link selection policy with CSI feedback can be summarized as Algorithm~\ref{algorithm:with_feedback}.

\begin{algorithm}[!ht]
\caption{Link Selection Policy with CSI Feedback}
\begin{algorithmic}[1]
\REQUIRE ~~\\
Instantaneous CSIs of two legitimate link, thresholds $\displaystyle \alpha$ and $\beta$ ($\alpha \geq 2^{R_s}-1$ and $\beta \geq 2^{R_s}-1$);
\ENSURE ~~\\
Link decision indicator $\displaystyle I_k$, $k \in \{1,2,\cdots,T\}$;
\FOR{ $\displaystyle k=1$; $k \leq T$; $k++$ }
  \STATE Calculate $\displaystyle \gamma_{a,r}[k]$ and $\displaystyle \gamma_{r,b}[k]$ based on the instantaneous CSIs;
  \IF{ $\displaystyle \gamma_{a,r}[k] \geq \alpha \land \frac{\gamma_{a,r}[k]}{\alpha} \geq \frac{\gamma_{r,b}[k]}{\beta}$ }
    \STATE $\displaystyle I_k=0$;
  \ELSIF{ $\displaystyle \gamma_{r,b}[k] \geq \beta \land \frac{\gamma_{r,b}[k]}{\beta} > \frac{\gamma_{a,r}[k]}{\alpha}$ }
    \STATE $\displaystyle I_k=1$;
  \ELSE
    \STATE $\displaystyle I_k=-1$;
  \ENDIF
\ENDFOR
\end{algorithmic}
\label{algorithm:with_feedback}
\end{algorithm}

\subsection{Link Selection Policy without CSI Feedback} \label{section:policy_without_CSI}

Regarding another case that Relay will not feed back the instantaneous CSI to Alice if Alice-to-Relay link is selected due to the system complexity concern, instead of adaptively adjusting the codeword rate to be the channel capacity, Alice uses a fixed rate $R_{a}$ to transmit the codewords. Considering that $R_a$ may be larger than the channel capacity which will incur the channel outage if Alice conducts the transmission, in order to avoid the channel outage, Relay should not select Alice-to-Relay link when it finds that $\gamma_{a,r}[k] < 2^{R_a}-1$ (even if $\gamma_{a,r}[k] \geq \alpha$). Therefore, our link selection policy without CSI feedback can be summarized as Algorithm~\ref{algorithm:without_feedback}.

\begin{algorithm}[!t]
\caption{Link Selection Policy without CSI Feedback}
\begin{algorithmic}[1]
\REQUIRE ~~\\
Instantaneous CSIs of two legitimate link, thresholds $\alpha$ and $\beta$ ($\alpha \geq 2^{R_s}-1$ and $\beta \geq 2^{R_s}-1$);
\ENSURE ~~\\
Link decision indicator $\displaystyle I_k$, $k \in \{1,2,\cdots,T\}$;
\FOR{ $\displaystyle k=1$; $k \leq T$; $k++$ }
  \STATE Calculate $\displaystyle \gamma_{a,r}[k]$ and $\displaystyle \gamma_{r,b}[k]$ based on the instantaneous CSIs;
  \IF{ $\displaystyle \gamma_{a,r}[k]\geq \max\{\alpha,2^{R_a}-1\}$ }
    \IF{ $\displaystyle \frac{\gamma_{a,r}[k]}{\alpha} \geq \frac{\gamma_{r,b}[k]}{\beta}$ }
      \STATE $\displaystyle I_k=0$;
    \ELSE
      \STATE $\displaystyle I_k=1$;
    \ENDIF
  \ELSIF{ $\displaystyle \gamma_{r,b}[k] \geq \beta$ }
    \STATE $\displaystyle I_k=1$;
  \ELSE
    \STATE $I_k=-1$;
  \ENDIF
\ENDFOR
\end{algorithmic}
\label{algorithm:without_feedback}
\end{algorithm}

\begin{figure}[t]
\centering
\includegraphics[width=1\linewidth]{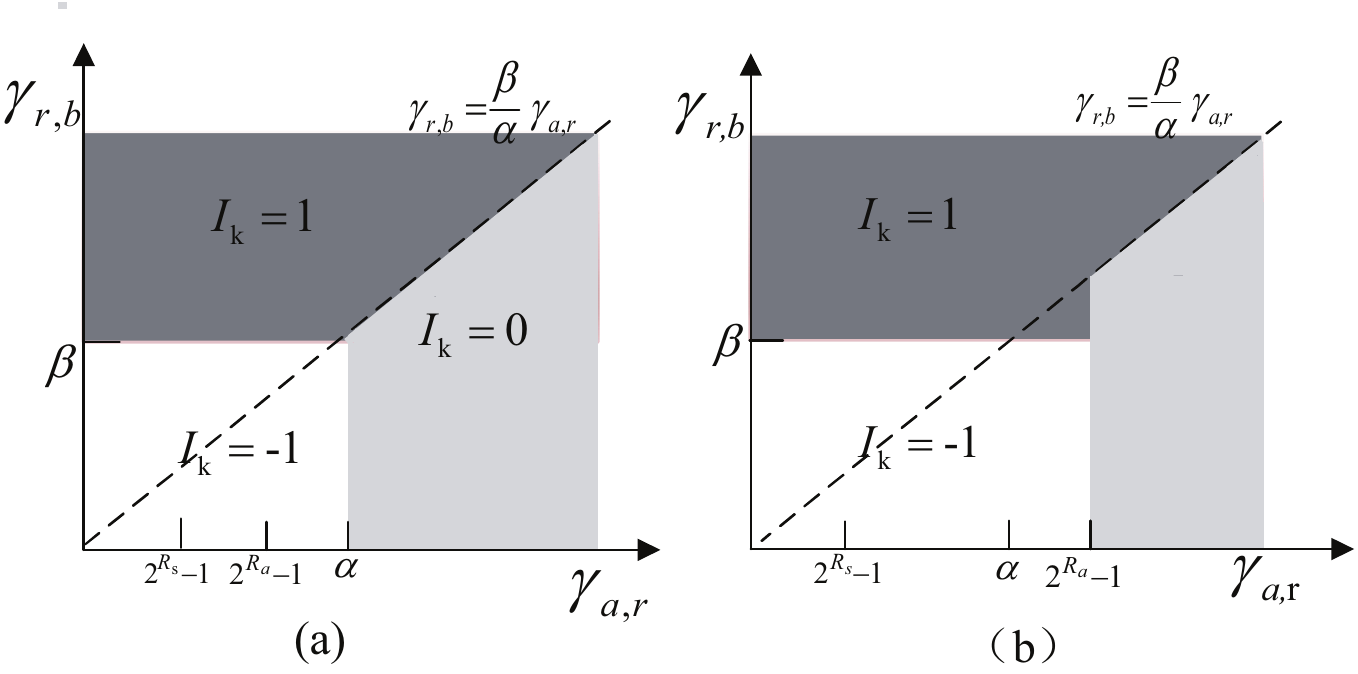}
\caption{The value of $I_k$ in different SNR regions. (a) $\alpha \geq 2^{R_a}-1$. (b) $\alpha<2^{R_a}-1$.}
\label{fig:indicator_value}
\end{figure}

In order to make a better understanding of our link selection policy, we illustrate in Fig.~\ref{fig:indicator_value} the value of $I_k$ in different SNR regions. We can see from Fig.~\ref{fig:indicator_value}(a) that when we set the threshold $\displaystyle \alpha \geq 2^{R_a}-1$, the value of $I_k$ in different SNR regions decided by the policy without CSI feedback is the same as that decided by the policy with CSI feedback. However, if we set the threshold $\displaystyle \alpha < 2^{R_a}-1$, when $\displaystyle 2^{R_a}-1 > \gamma_{a,r}[k] > \alpha$, $\gamma_{r,b}[k] > \beta$ and $\frac{\gamma_{a,r}[k]}{\alpha} \geq \frac{\gamma_{r,b}[k]}{\beta}$, Relay-to-Bob link will also be selected to transmit message (i.e., $I_k=1$), as shown in the triangle area of Fig.~\ref{fig:indicator_value}(b).

\section{General Optimization Problem Formulation} \label{section:problem_formulation}
  \par In this section, we first derive the general expression for secrecy outage capacity, end-to-end secrecy outage probability and exact secrecy throughput,  then we formulate the general optimization problems about these performance metrics.

\subsection{Performance Metrics}
\subsubsection{Secrecy Outage Probability}
   \par Differing form the tradition SOP \cite{08TIT:WirelessInformation-TheoreticSecurity} \cite{13JSAC:SOP},
   a better expression for SOP \cite{13rethink:SOP} is defined as the conditional probability that capacity of the wiretap channels exceeds the rate redundancy under message successful transmission (i.e., $\mathcal{P}_{so}=\text{Pr}[{C_s}^+\!\!<\!R_s\!|\text{ message transmission}]$, where ${C_s}^+=\text{max}\{0,R_{a,r}-C_e\}$ ). Expediently, we use $\Gamma_{a,r}[k]$ and $\Gamma_{r,b}[k]$  as secrecy outage variables in two hops respectively, which are expressed as
\begin{eqnarray}
&\Gamma_{a,r}[k]=
\begin{cases}
1, & {R_{s} > R_{a,r}[k]-C_{a,e}[k]}\\
0, & \text{otherwise}
\end{cases} \\
\label{equ:10}
&\Gamma_{r,b}[k]=
\begin{cases}
1, & {R_{s} > R_{r,b}[k]- C_{r,e}[k]}\\
0, & \text{otherwise} .
\end{cases}
\label{eq:SOP variables}
\end{eqnarray}

  Considering there exists two wiretap channels, the system SOP can be characterized as a combination of the individual SOP of each hop, and end-to-end SOP is expressed as
\begin{align}
\boldsymbol{\rho}_{so}\triangleq1-(1-{\mathcal{P}^1_{so}})(1-{\mathcal{P}^2_{so}}),
\label{equ:12}
\end{align}
  where ${\mathcal{P}^1_{so}}\!\triangleq\!\text{Pr}[\Gamma_{a,r}[k]=1|I_k=0]$ and ${\mathcal{P}^2_{so}}\!\triangleq\!\text{Pr}[\Gamma_{r,b}[k]=1|I_k=1]$, which are derived in Section \ref{section:performance_evaluation}.

\subsubsection{Secrecy Outage Capacity}

  \par The secrecy outage constraint throughput (SOCT) is defined as the largest achievable secrecy rate under end-to-end SOP constraint at destination
\begin{equation}
\begin{split}
\varPhi&=\text{maximum}\{\mathbb{E}_{I_k}[\Theta{(R_s,I_k)}]\} \\
&\text{s.t.}\quad  \boldsymbol{\rho}_{so}\le \nu,
\label{equ:13}
\end{split}
\end{equation}
  where $\Theta{(R_s,I_k)}=R_s$ when $I_k=1$ and $\Theta{(R_s,I_k)}=0$ otherwise. Furtherly, the SOCT can be generally rewritten as
\begin{equation}
\begin{split}
{\varPhi}=\text{maximum}&\{\lim\limits_{N\rightarrow{\infty}}\frac{1}{N} \sum_{k=1}^N (|I_k+\frac{1}{2}|-\frac{1}{2})R_s\} \\
\quad &\text{s.t.}\quad   \boldsymbol{\rho}_{so}\le \nu.
\label{equ:14}
\end{split}
\end{equation}

\subsubsection{Secrecy Throughput}
  \par 
  The average rate securely and reliably delivered to the queue of Bob via Relay over multiple scheduling is defined as the end-to-end secrecy throughput. A novel formulation (called successful transmission probability $\boldsymbol{\rho}_{srt}$ in \cite{16TVT:STP}) to characterize the reliability and security levels of transmission in both two hops oh the buffer-aided relaying system, which are defined as
\begin{align}
\boldsymbol{\rho}^1_{srt}=\text{Pr}\{R_{a,r}\!\le\!C_{a,r},R_{s}\!\le\!R_{a,r}\!-\! C_{a,e}|I_k=0\}
\label{equ:15}
\end{align}
  and
\begin{align}
\boldsymbol{\rho}^2_{srt}=\text{Pr}\{R_{r,b}\!\le\!C_{r,b},R_{s}\!\le\!R_{r,b}\!-\! C_{r,e}|I_k=1\}.
\label{equ:16}
\end{align}
  \par Using the notation introduced above, the general expression for the average successful arriving rate $\tau_{a,r}$ in the buffer queue of Relay is derived as
\begin{align}
\tau_{a,r}&\triangleq \boldsymbol{\rho}^1_{srt}\text{Pr}[I_k=0]\text{min}(R_s,R_{a,r}[k])  \nonumber \\
&\overset{\Xi}{=}\lim\limits_{N\rightarrow{\infty}}\frac{1}{N} \sum_{k=1}^N(1-\Gamma_{a,r}[k])(1-|I_k|)R_s
\label{equ:17}
\end{align}
where $\Xi$ is due to $R_s<R_{a,r}[k]$ always satisfies when $\Gamma_{a,r}[k]=0$.
Similarly, the average successful accepting rate $\tau_{r,b}$ (i.e., the secrecy throughput) securely successfully delivered to Bob can be generally expressed as
\begin{align}
\tau_{r,b}&\triangleq \boldsymbol{\rho}^2_{srt}\text{Pr}[I_k=1]\text {min}(R_{s},Q[k-1])  \nonumber \\
&=\lim\limits_{N\rightarrow{\infty}}\frac{1}{N}\sum_{k=1}^N(1-\Gamma_{r,b}[k])(|I_k+\frac{1}{2}|-\frac{1}{2})R^{'}_s,
\label{equ:18}
\end{align}
  where $R^{'}_s=\text {min}(R_{s},Q[k-1])$.
  \par Note that $\tau_{a,r}\ge\tau_{r,b}$ is valid because of the buffer-aided relaying protocol and when $\tau_{a,r}>\tau_{r,b}$, the queue of buffer is said to be unstable or in the absorbing sate according to \cite{13JSAC:SOP}. In our system, we have following lemma:
  \begin{lemma}
  \label{lemma 1} When the secrecy throughput $\tau_{r,b}$ reached the maximum, the necessary condition for optimal link selection policy is that the buffer queue is always at the edge of non-absorbing (i.e., $\tau_{a,r}=\tau_{r,b}$) and the link selection parameters $\alpha$, $\beta$ are optimal.
  \end{lemma}
  \begin{IEEEproof}\,\,Because the \ref{lemma 1} only is an extension of theorem 1 in \cite{13JSAC:SOP} in our network scene with Relay having three working states (i.e., accepting, transmitting and idle), we just provide a brief proof. We denote the set of indices with $I_k >0$ by $M$ and the set of  $I_k \le 0$ by $\bar M$ (where $\bar A$ is the complementary set of set $A$). Assume that we have a link selection policy resulting in $\tau_{a,r}> \tau_{r,b}$ (i.e., the queue is in the absorbing state). The policy can always can be improved by moving the indices $k$ in $\bar M$ to $M$ to to increase the secrecy throughput until $\tau_{a,r}= \tau_{r,b}$ (i.e., the queue is at the edge of non-absorbing), which leads to an increase of $\tau_{r,b}$ at the expense of a decrease of $\tau_{a,r}$. From the law of the conservation of flow, we know that both $\tau_{a,r}$ and $\tau_{r,b}$  will decrease if we move the indices $\alpha$ and $\beta$ further once the point $\tau_{a,r}= \tau_{r,b}$ is reached. Therefore, When the buffer queue switches form absorbing state to non-absorbing state, the secrecy throughput exactly reached the maximum value.
  \end{IEEEproof}
  \begin{lemma}
  \label{lemma 2}
   When the link selection policy is optimal, the event $R_s>Q[k-1]$ can be negligible and the maximum exact secrecy throughput can be generally written as
 \begin{align}
\tau_{r,b}\triangleq \lim\limits_{N\rightarrow{\infty}}\frac{1}{N} \sum_{k=1}^N(1-\Gamma_{r,b}[k])(|I_k+\frac{1}{2}|-\frac{1}{2})R_s.
\end{align}
\end{lemma}
  \begin{IEEEproof} \,\,\,We denote the set of indices $k$ with $R_s\le Q_r[k-1]$ by $C$ and the set of  $R_s> Q_r[k-1]$ by $\bar C$. we also denote the cardinalities  of $C$ and $\bar C$ as $N-|\bar C|$ and $|\bar C|$, respectively. Thus, if the queue is absorbing, $R_s\le Q_r[k-1]$ always holds, which means $|\bar C|/N \rightarrow 0$ $(N \rightarrow \infty )$. As a result, the secrecy outage capacity is
\begin{equation*}
\begin{split}
\tau_{r,b}&\triangleq \lim\limits_{N\rightarrow{\infty}}\frac{1}{N}\sum_{k=1}^N(1-\Gamma_{r,b}[k])(|I_k+\frac{1}{2}|-\frac{1}{2})R^{'}_s\\
&=\lim\limits_{N\rightarrow{\infty}}\frac{1}{N}(\sum_{k\in C}(1-\Gamma_{r,b}[k])(|I_k+\frac{1}{2}|-\frac{1}{2})R_{s} \\
&\,\,\,+\sum_{k\in \bar C}(1-\Gamma_{r,b}[k])(|I_k+\frac{1}{2}|-\frac{1}{2})Q[k-1]) \\
&=\lim\limits_{N\rightarrow{\infty}}\frac{N-|\bar C|}{N} \sum_{k\in  C}(1-\Gamma_{r,b}[k])(|I_k+\frac{1}{2}|-\frac{1}{2})R_{\!s}\\
&=\lim\limits_{N\rightarrow{\infty}}\frac{1}{N} \sum_{k=1}^N(1-\Gamma_{r,b}[k])(|I_k+\frac{1}{2}|-\frac{1}{2})R_s.
\end{split}
\end{equation*}
\end{IEEEproof}

\subsection{General Optimal Formulation}
  \par Now we are ready to formulate the general optimal problem. We respectively derive the optimal link selection policy (i.e., optimal $\alpha$ and $\beta$) and optimal secrecy rate $R_{s}$ that the encoders use for encoding to maximize secrecy outage capacity under a desired certain outage probability, and minimize SOP under the desired certain secrecy outage capacity.
  \par The general problem for maximizing the secrecy outage capacity can be formulated as
\begin{equation*}
\begin{split}
 \mathop{\text{P1:}\,\,\text{max}}\limits_{\,\,\,\,\,\,\,\alpha,\beta,R_{s}} \;\;\;\; &\lim\limits_{N\rightarrow{\infty}}\frac{1}{N} \sum_{k=1}^N (|I_k+\frac{1}{2}|-\frac{1}{2})R_s\\
s.t. \;\;\;\;&\text{C1}:\boldsymbol{\rho}_{id} \le \mu \\
                          &\text{C2}:\boldsymbol{\rho}_{so}=1-(1-{\mathcal{P}^1_{so}})(1-{\mathcal{P}^2_{so}}) \le \nu   \\
                          &\text{C3}:R_{s}>0,\alpha>0,\beta>0.
\end{split}
\end{equation*}
  Because the high probability that Relay keeps idle (i.e., $\boldsymbol{\rho}_{id}= \text{Pr}[I_k=-1]$) leads to the higher delay, constraint C1 is required. In C2, $\nu$ denotes the desired SOP of the system. Note that C1 and C2  represent two QoS metrics for information security and delay in cooperative communication, respectively.
  \par Then the optimal problem to minimum the end-to-end SOP which represents transmission security performance of the system under secrecy outage capacity constraints, is generally formulated as following

 \begin{equation*}
\begin{split}
 \mathop{\text{P2:}\,\,\text{min}}\limits_{\,\,\,\,\,\,\,\alpha,\beta,R_{s}} \;\;\;\; &\boldsymbol{\rho}_{so}=1-(1-{\mathcal{P}^1_{so}})(1-{\mathcal{P}^2_{so}})\\
s.t. \;\;\;\;&\text{C1}:\boldsymbol{\rho}_{id} \le \mu \\
                          &\text{C2}:\lim\limits_{N\rightarrow{\infty}}\frac{1}{N} \sum_{k=1}^N (|I_k+\frac{1}{2}|-\frac{1}{2})R_s \!\ge\!\xi   \\
                          &\text{C3}:R_{s}>0,\alpha>0,\beta>0,
\end{split}
\end{equation*}
  where $\theta$ is the desired minimum secrecy outage capacity.
\par Based on \emph{Lemma 1} and \emph{Lemma 2}, we formulate the optimal problem about exact secrecy throughput as follow

 \begin{equation*}
\begin{split}
 \mathop{\text{P3:}\,\,\text{max}}\limits_{\,\,\,\,\,\,\,\alpha,\beta,R_{s}} \tau_{r,b}= &\lim\limits_{N\rightarrow{\infty}}\frac{1}{N} \sum_{k=1}^N(1-\Gamma_{r,b}[k])(|I_k+\frac{1}{2}|-\frac{1}{2})R_s\\
s.t. \;\;\;\;&\text{C1}:\boldsymbol{\rho}_{id} \le \mu \\
                          &\text{C2}:\tau_{a,r}(\alpha,\beta,R_{s})=\tau_{r,b}(\alpha,\beta,R_{s}) \\
                          &\text{C3}:R_{s}>0,\alpha>0,\beta>0,
\end{split}
\end{equation*}
  where constraint C2 ensures that the link selection policy is optimal to make the exact secrecy throughput reach maximum value.

\begin{figure*}[b]
  \hrulefill
\begin{align*}
{\mathcal{P}^1_{so}}\triangleq\left\{\frac{2^{R_s}\bar{r}_{a,e}}{2^{R_s}\bar{r}_{a,e}+\bar{r}_{a,r}}\text{exp}\left(-{\frac{\alpha}{\bar{r}_{a,r}}-\frac{\alpha+1-2^{R_s}}{\bar{r}_{a,e}2^{R_s}}}\right)
\left[1-\frac{\frac{\bar{r}_{r,b}}{\bar{r}_{a,r}}+\frac{\bar{r}_{r,b}}{\bar{r}_{a,e}2^{R_s}}}{\frac{\beta}{\alpha}+\frac{\bar{r}_{r,b}}
{\bar{r}_{a,r}}+\frac{\bar{r}_{r,b}}{\bar{r}_{a,e}2^{R_s}}}\text{exp}\left(-{\frac{\beta}{\bar{r}_{r,b}}}\right)\right]\right\}/\boldsymbol{\rho}_{a}
\tag{27}
\end{align*}
\begin{align*}
{\mathcal{P}^2_{so}}\triangleq\left\{\frac{2^{R_s}\bar{r}_{r,e}}{2^{R_s}\bar{r}_{r,e}+\bar{r}_{r,b}}\text{exp}\left(-{\frac{\beta}{\bar{r}_{r,b}}-\frac{\beta+1-2^{R_s}}{\bar{r}_{r,e}2^{R_s}}}\right)
\left[1-\frac{\frac{\bar{r}_{a,r}}{\bar{r}_{r,b}}+\frac{\bar{r}_{a,r}}{\bar{r}_{r,e}2^{R_s}}}{\frac{\alpha}{\beta}+\frac{\bar{r}_{a,r}}
{\bar{r}_{r,b}}+\frac{\bar{r}_{a,r}}{\bar{r}_{r,e}2^{R_s}}}\text{exp}\left(-{\frac{\alpha}{\bar{r}_{a,r}}}\right)\right]\right\}/\boldsymbol{\rho}_{r}
\tag{28}
\end{align*}
\begin{align*}
\tau_{a,r}=
\text{exp}\left(-{\frac{\alpha}{\bar{r}_{a,r}}}\right)\left[R_s-\frac{\alpha \bar{r}_{r,b}R_s\text{exp}\left(-{\frac{\beta}{\bar{r}_{r,b}}}\right)}{\alpha\bar{r}_{r,b}+\beta\bar{r}_{a,r}}\right]
+\frac{R_s\bar{r}_{a,e}2^{R_s}}{2^{R_s}\bar{r}_{a,e}+\bar{r}_{a,r}}\text{exp}\left(-\frac{\alpha}{\bar{r}_{a,r}}-\frac{\alpha+1-2^{R_s}}{\bar{r}_{a,e}2^{R_s}}\right)
\left[1-\frac{(\frac{\bar{r}_{r,b}}{\bar{r}_{a,r}}+\frac{\bar{r}_{r,b}}{2^{R_s}\bar{r}_{a,e}})e^{-{\frac{\beta}{\bar{r}_{r,b}}}}}
{\frac{\beta}{\alpha}+\frac{\bar{r}_{r,b}}{\bar{r}_{a,r}}+\frac{\bar{r}_{r,b}}{2^{R_s}\bar{r}_{a,e}}}  \right]
\tag{29}
\end{align*}
\begin{align*}
\tau_{r,b}= \text{exp}\left(-{\frac{\beta}{\bar{r}_{r,b}}}\right)(R_s-R_s\frac{\beta\bar{r}_{a,r}e^{-{\frac{\alpha}{\bar{r}_{a,r}}}}}{\beta\bar{r}_{a,r}+\alpha\bar{r}_{r,b}})
+\frac{R_s\bar{r}_{r,e}2^{R_s}}{2^{R_s}\bar{r}_{r,e}\,+\,\bar{r}_{r,b}}\,e^{-(\frac{\beta}{\bar{r}_{r,b}}+\frac{\beta+1-2^{R_s}}{\bar{r}_{r,e}2^{R_s}})}
\left[1-\frac{(\frac{\beta\bar{r}_{a,r}}{\alpha\bar{r}_{r,b}}+\frac{\beta\bar{r}_{a,r}}{\alpha2^{R_s}\bar{r}_{r,e}})\,e^{-{\frac{\alpha}{\bar{r}_{a,r}}}}}
{1+\frac{\beta\bar{r}_{a,r}}{\alpha\bar{r}_{r,b}}+\frac{\beta\bar{r}_{a,r}}{\alpha2^{R_s}\bar{r}_{r,e}}}  \right]
\tag{30}
\end{align*}
	\hrulefill
	\vspace*{4pt}
\end{figure*}

\section{PERFORMANCE ANALYSIS AND OPTIMIZATION} \label{section:performance_evaluation}
  \par In this section, we first derive the closed-form expressions for above introduced performance parameters. Based on these closed-form expressions, we re-formulate the detail optimal problem P1 and P2 under adaptive-rate and fixed-rate transmission models, respectively.

\subsection{Performance Analysis For Adaptive-rate Transmission Model}
  \par According the link selection scheme (9), the transmission probability at time slot $k$ when Alice is selected to transmit message can be given by

\begin{align}
\boldsymbol{\rho}_a(\alpha,\beta) &= \text{Pr}[\gamma_{a,r}[k]\ge \text{max}\{\alpha,\frac{\alpha}{\beta}\gamma_{r,b}[k]\}] \nonumber\\
 &=\text{Pr}[\gamma_{a,r}[k]\ge\frac{\alpha}{\beta}\gamma_{r,b}[k],\gamma_{r,b}[k]\ge\beta] \nonumber\\
 &\;\;\;\;\;+\text{Pr}[\gamma_{a,r}[k]\ge\alpha,\gamma_{r,b}[k]\le\beta] \nonumber\\
&=\text{exp}\left(-\frac{\alpha}{\bar \gamma_{a,r}}\right)-\frac{\alpha\bar{\gamma}_{r,b}\text{exp}\left(-\frac{\alpha}{\bar{\gamma}_{a,r}}-\frac{\beta}{\bar{\gamma}_{r,b}}\right)}{\alpha\bar{\gamma}_{r,b}+\beta\bar{\gamma}_{a,r}}. \end{align}
  By symmetry, we can obtain the transmission probability for Relay

\begin{align}
\boldsymbol{\rho}_r(\alpha,\beta) &= \text{Pr}[\gamma_{r,b}[k]\ge \text{max}\{\beta,\frac{\beta}{\alpha}\gamma_{a,r}[k]\}] \nonumber\\
&=\text{exp}\left(-\frac{\beta}{\bar{\gamma}_{r,b}}\right)-\frac{\beta\bar{\gamma}_{a,r}\text{exp}\left(-\frac{\beta}{\bar{\gamma}_{r,b}}-\frac{\alpha}{\bar{\gamma}_{a,r}}\right)}{\beta\bar{\gamma}_{a,r}+\alpha\bar{\gamma}_{r,b}}. \end{align}
So the probability that Relay is idle due to security consideration can be presented as

\begin{align}
\boldsymbol{\rho}_{id}(\alpha,\beta) &=1-\boldsymbol{\rho}_a(\alpha,\beta)-\boldsymbol{\rho}_r(\alpha,\beta) \nonumber\\
&=\left[1-\text{exp}\left(-\frac{\alpha}{\bar{\gamma}_{a,r}}\right)\right]\!\left[1-\text{exp}\left(-\frac{\beta}{\bar{\gamma}_{r,b}}\right)\right].
\end{align}
\par \emph{Remark 2:}\,\,\,\,  According to the link selection scheme (9), the condition $\text{min}\{\alpha, \beta\}> 2^{R_s}-1$ is required to avoid the decoding outage, i.e., $R_s>R_{a,r}$.
\par Then we derive the E2E secrecy outage probability. Based on \uppercase\expandafter{\romannumeral4}, the SOP in the first hop can be expressed as

\begin{align}
{\mathcal{P}^1_{so}}&\triangleq   \text{Pr}[C_{a,r}-C_{a,e}<R_s|I_k=0] \nonumber \\
&=\frac{\text{Pr}[\text{max}\{\alpha,\frac{\alpha}{\beta}\gamma_{r,b}\}<\gamma_{a,r}<2^{R_s}-1]}{\boldsymbol{\rho}_{a}(\alpha,\beta)}.
\end{align}
  Similarly, the SOP in the second hop is

\begin{align}
{\mathcal{P}^1_{so}}&\triangleq \text{Pr}[C_{r,b}-C_{r,e}<R_s|I_k=1] \nonumber \\
&=\frac{\text{Pr}[\text{max}\{\beta,\frac{\beta}{\alpha}\gamma_{a,r}\}<\gamma_{r,b}<2^{R_s}-1]}{{\boldsymbol{\rho}_{r}(\alpha,\beta)}}
\end{align}
  substituting the p.d.f.s of $\gamma_{a,r}$,$\gamma_{r,b}$,$\gamma_{a,e}$ and $\gamma_{a,r}$, the detailed expression of $\mathcal{P}^1_{so}$ and $\mathcal{P}^2_{so}$ can be derived as (27)(28), which are shown at the bottom of this page, respectively. Thus the E2E SOP can be obtained based on (14).
  \par Now we are ready to reformulate the  optimal problems about secrecy capacity and SOP. We denote the secrecy outage capacity maximization and SOP minimization as PA1 and PA2 in adaptive-rate transmission model, respectively.
  \par Problem PA1 is formulated as

\begin{equation*}
\label{MAX_SOCT}
\begin{split}
 \mathop{\text{PA1:}\,\,\text{max}}\limits_{\,\,\,\,\,\,\,\alpha,\beta,R_s} \;\;\;\; &\varPhi(\alpha,\beta,R_s)\\
\text{subject to} \;\;\;\;&\text{C1}:\boldsymbol{\rho}_{so}(\alpha,\beta,R_s) \le \mu \\
                          &\text{C2}:\boldsymbol{\rho}_{id}(\alpha,\beta) \le \nu   \\
                          &\text{C3}:R_s>0,\text{min}\{\alpha,\beta\}>2^{R_s}-1,
\end{split}
\end{equation*}
  and for Problem PA2, we have

 \begin{equation*}
\begin{split}
 \mathop{\text{PA2:}\,\,\text{min}}\limits_{\,\,\,\,\,\,\,\alpha,\beta,R_s} \;\;\;\; &P_{sop}(\alpha,\beta,R_s)\\
\text{subject to} \;\;\;\;&\text{C1}:\varPhi(\alpha,\beta,R_s) \ge \theta \\
                          &\text{C2}:P_{ip}(\alpha,\beta) \le \nu   \\
                          &\text{C3}:R_s>0,\text{min}\{\alpha,\beta\}>2^{R_s}-1,
\end{split}
\end{equation*}
  \par It is notable that (27) and (28) include the transcendental function and it's hard to analyze the monotonicity of $\boldsymbol{\rho}_{so}$, thus the closed-form solutions in PA1 and PA2 are generally not possible. Inspired by the Zoutendijk Method \cite{1999:nonlinear}, we apply the algorithm \ref{algorithm:Improved Feasible Direction Method} to solve the optimal problems. before introducing the algorithm \ref{algorithm:Improved Feasible Direction Method}, we introduce the the following lemma:
\begin{lemma}
\label{lemma 3}
  Suppose the feasible point $x^{k}=(\alpha^{(k)},\beta^{(k)},{R_s}^{(k)})$ in problem \ref{MAX_SOCT} has obtained after $k$-th iteration, finding the strictly feasible descent direction $d$ at the point is equivalent to the solution the following linear program problem:
\begin{align}
\label{LP problem}
&\;\;\;\;\;\;\;\;\;\;\;\;\;\min \limits_{d,\delta} \delta \nonumber \\
&s.t.
\begin{cases}
&-d^{\small{T}}\nabla \varPhi(x^{(k)})\le \delta  \\
&d^{\small{T}}\nabla -g_i(x^{(k)})\le\delta  \\
&\mid d_j\mid < 0, j=1,2,3 \\
\end{cases}
\end{align}
\end{lemma}
  where $g_1(x^{(k)})=\mu-\boldsymbol{\rho}_{so}(\alpha^{(k)},\beta^{(k)},R_s^{(k)})$, $g_2(x^{(k)})=\nu-\boldsymbol{\rho}_{id}(\alpha^{(k)},\beta^{(k)})$, $g_3(x^{(k)})=(\text{min}(\alpha^{(k)},\beta^{(k)})-2^{R_s^{(k)}}+1)$ and $\nabla g_i(x^{(k)})$ is the first derivative of function $g_i(x^{(k)})$ with respect to $x^{(k)}$.
\begin{proof}
  According to \cite{1967convergence}\cite{04:convex}, $d$ is the decline direction of $-\varPhi(x^{(k)})$\footnote{the equivalent problem of maximizing $\varPhi(x^{(k)})$ is the one of minimizing $-\varPhi(x^{(k)})$.Thus this paper focuses on the the feasible decline direction of $-\varPhi(x^{(k)})$.} if and only if $-d^{\small{T}}\nabla\varPhi(x^{(k)})<0 $ at the point $x^{(k)}$. Furthermore, if $d^{\small{T}}\nabla g_i(x^{(k)})>0$ holds at $x^{(k)}$, $d$ is called the strictly feasible direction. Thus we call $d$ as the strictly feasible descent direction, if there exists $\delta$ such that
\begin{align}
\label{feasible descent direction}
\begin{cases}
&-d^{\small{T}}\nabla \varPhi(x^{(k)})\le \delta  \\
&-d^{\small{T}}\nabla g_i(x^{(k)})\le\delta  \\
&\delta<0.
\end{cases}
\end{align}
  \par In order to find a feasible descent direction at $x^{(k)}$, only need to find out $d$ and the minimum value of $\delta$ which hold the condition (\ref{feasible descent direction}). $\mid d_j\mid \le 1$ is added to guarantee a finite optimal solution. Thus, finding a feasible descent direction formulates as the linear program problem (\ref{LP problem}).
\end{proof}
  The Algorithm \ref{algorithm:Improved Feasible Direction Method} is as follows:

\begin{algorithm}[!ht]
\caption{The Improved Feasible Direction Method}
\begin{algorithmic}[1]
\renewcommand{\algorithmicrequire}{\textbf{Initialization:}}
\REQUIRE ~~\\
 the initial feasible point $x^{(0)}$, $\epsilon_0>0$ and the maximum tolerance of objective function $\varepsilon>0$, $0\Longrightarrow k$ ;
\ ~~\\
\ENSURE ~~\\
 The optimal link selection parameters $x^*$ \\
 \STATE Step 1: Determine the effective constraint indicator set:
   $I(x^{(k)},\epsilon_k)=\{\mid i\mid0\le g_i(x^{(k)})\le\epsilon_k, 1\le i\le3\}$, then compute $-\nabla \varPhi(x^{(k)})$;  \\
  \IF{ $I(x^{(k)},\epsilon_k)=\varnothing$ and $\parallel -\nabla \varPhi(x^{(k)}) \parallel\le\varepsilon$}
    \STATE stop iteration and $x^*=x^{(k)}$;
  \ELSIF{ $\parallel -\nabla \varPhi(x^{(k)}) \parallel>\varepsilon$}
    \STATE set $\nabla \varPhi(x^{(k)}=d^{(k)}, \delta^{(k)}=-1$ goto Step 3;
  \ELSE
    \STATE goto Step 2; \\
  \ENDIF
  \STATE Step 2: Compute the linear programming problem (\ref{LP problem}), then return $d^{(k)}, \delta_k$ ;
  \IF {$\delta_k=0$ and $\epsilon_k<\varepsilon$}
    \STATE stop iteration and $x^*=x^{(k)}$;
  \ELSE
   \STATE {update $\epsilon_k=\frac{\epsilon_k}{2}$, goto Step 1}; \\
  \ENDIF
  \STATE Step 3: One-dimensional search with direction $d^{(k)}$, For $i\notin I(x^{(k)},\epsilon_k)$, compute $a_{max}=\text{min}\{ t_i\vert(g_i(x^{(k)})+t_id^{(k)})=0,t_i>0\}$, then obtain $a_k$ by computing
  \begin{align*}
  \begin{cases}
  &\text{min}-\varPhi(x^{(k)}+ad^{(k)})\\
  &0\le a\le a_{max}\\
  \end{cases}
  \end{align*}
  set $x^{(k+1)}= x^{(k)}+a_kd^{(k)}$;
  \IF {$\parallel x^{(k+1)}-x^{(k)}\parallel < \varepsilon$}
    \STATE {stop iteration and $x^*=x^{(k+1)}$;} \\
  \ELSE
  \STATE {update $\epsilon_k=\epsilon_k$ for $\epsilon_k\le-\delta _k$, $\epsilon_k=\frac{\epsilon_k}{2}$ for $\epsilon_k>-\delta _k, k=k+1$, goto Step 1.}
   \ENDIF
\end{algorithmic}
\label{algorithm:Improved Feasible Direction Method}
\end{algorithm}

  \par Because of the  non-negative, ergodic and stationary process, we have the following proposition.
  \par \emph{Proposition 1:} \,\,When Alice transmits message to Relay with adaptive rate, the average accept rate $\tau_{a,r}$ can be expressed as

\begin{align*}
\tau_{a,r}=\mathrm{E}\{(1-\Gamma_{a,r}[k])(1-|I_k|)R_s\}
\end{align*}
  and the detail expression is given in (29).
  \par \emph{Proof:}\,\, See Appendix A.
  \par \emph{Proposition 2:}\, From the above proof process of Lemma 2, we note the influence of the event $R_s> Q_r[k-1]$ is negligible. Therefore, when the link
  selection strategy is optimal to maximize the secrecy outage capacity by optimizing the scalars $\alpha$ and $\beta$. the secrecy outage capacity can be given  as

\begin{align*}
\tau_{r,b}&=\mathrm{E}\{(1-\Gamma_{r,b}[k])(|\frac{1}{2}+I_k|-\frac{1}{2})R_s\}
\end{align*}
  and the detail expression is given in (30).
   \par \emph{Proof:}\,\, See Appendix A.
   Thus the optimal problem PA3 of maximizing the  exact secrecy throughput formulates as

 \begin{equation*}
\begin{split}
 \text{PA3:}\mathop{\,\,\text{maximize}}\limits_{\alpha,\beta,R_s} \;\;\;\; &\tau_{r,b}(\alpha,\beta,R_s))\\
\text{subject to} \;\;\;\;&\text{C2}:\boldsymbol{\rho}_{id}(\alpha,\beta) \le \nu   \\
                          &\text{C3}:\tau_{a,r}(\alpha,\beta,R_{sec})=\tau_{r,b}(\alpha,\beta,R_s) \\
                          &\text{C4}:0<R_s<R_{a}\\
                          &\text{C5}:\text{min}(\alpha,\beta)\ge2^{R_s}-1
\end{split}
\end{equation*}

\begin{figure*}[b]
  \hrulefill
   \begin{align}
\mathit{P}^*_r(\alpha,\beta)&= \text{Pr}[\gamma_{r,b}[k]\ge max\{\beta,\frac{\beta}{\alpha}\gamma_{a,r}[k]\}]+\text{Pr}[\alpha<\gamma_{a,r}[k]<2^{R_a}-1,\beta<\gamma_{r,b}[k]<\frac{\beta}{\alpha}\gamma_{a,r}[k]]\nonumber\\
&=\begin{cases}
e^{-\frac{\beta}{\bar{\gamma}_{r,b}}}+\frac{\alpha\bar{\gamma}_{r,b}e^{-(\frac{2^{R_a}-1}{\bar{\gamma}_{a,r}}+\frac{\beta(2^{R_a}-1)}{\alpha\bar{\gamma}_{r,b}})}}{\alpha\bar{\gamma}_{r,b}+\beta\bar{\gamma}_{a,r}} -e^{-(\frac{\beta}{\bar{\gamma}_{r,b}}+\frac{2^{R_a}-1}{\bar{\gamma}_{a,r}})}&\alpha<2^{R_a}-1
\\
\text{the same as (27)}&\alpha\ge2^{R_a}-1
\end{cases}
\tag{30}
\end{align}
\begin{align}
{\mathcal{P}^{2*}_{sop}}&= \text{Pr}[C_{r,b}-C_{r,e}<R_s|(\,\gamma_{r,b}[k]\ge max\{\beta,\frac{\beta}{\alpha}\gamma_{a,r}[k]\}||
\alpha<\gamma_{a,r}[k]<2^{R_a}-1,\beta<\gamma_{r,b}[k]<\frac{\beta}{\alpha}\gamma_{a,r}[k]\,)] \nonumber\\
&=\left\{\mathcal{P}^2_{sop}P_r+\text{Pr}[C_{r,b}-C_{r,e}<R_s,\alpha<\gamma_{a,r}[k]<2^{R_a}-1,\beta<\gamma_{r,b}[k]<\frac{\beta}{\alpha}\gamma_{a,r}[k]]\right\}/P^*_r\nonumber\\
&=\frac{\mathcal{P}^2_{sop}P_r}{P^*_r}+\frac{\gamma_{r,e}2^{R_s}}{P^*_r(\gamma_{r,e}2^{R_s}+\gamma_{r,b})} \left[ \frac{\frac{\beta\gamma_{a,r}}{\alpha\gamma_{r,b}}+\frac{\beta\gamma_{a,r}}{\alpha2^{R_s}\gamma_{r,e}}}{1+\frac{\beta\gamma_{a,r}}{\alpha\gamma_{r,b}}+\frac{\beta\gamma_{a,r}}{\alpha2^{R_s}\gamma_{r,e}}}  e^{-({\frac{\alpha}{\gamma_{a,r}}+\frac{\beta}{\gamma_{r,b}}+\frac{\beta+1-2^{R_a}}{2^{R_s}\gamma_{r,e}}})}-e^{-({\frac{2^{R_a}-1}{\gamma_{a,r}}+\frac{\beta}{\gamma_{r,b}}+\frac{\beta+1-2^{R_a}}{2^{R_s}\gamma_{r,e}}}}) \right. \nonumber\\
&\,\,\,\,\,\,\left.-\frac{1}{1+\frac{\beta\gamma_{a,r}}{\alpha\gamma_{r,b}}+\frac{\beta\gamma_{a,r}}{\alpha2^{R_s}\gamma_{r,e}}}-e^{-({\frac{2^{R_a}-1}{\gamma_{a,r}}+\frac{\beta(2^{R_a}-1)}{\alpha\gamma_{r,b}}+\frac{\beta(2^{R_a}-1)+1-2^{R_a}}{\alpha2^{R_s}\gamma_{r,e}}})}\right]
\tag{32}
\end{align}
\begin{align}
\tau^*_{r,b}
&=\begin{cases}
e^{-\frac{\beta}{\bar\gamma_{r,b}}}\left[{\frac{\beta\bar\gamma_{a,r}}{\beta\bar\gamma_{a,r}+\alpha\bar\gamma_{r,b}}e^{-\frac{\alpha}{\bar\gamma_{a,r}}}}
-e^{-\frac{2^{R_a}-1}{\bar\gamma_{a,r}}}\right]+e^{-(\frac{2^{R_a}-1}{\bar\gamma_{a,r}}+\frac{\beta(2^{R_a}-1)}{\alpha\bar\gamma_{r,b}})}\left[\frac{\alpha\bar\gamma_{r,b}}{\alpha\bar\gamma_{r,b}
+\beta\bar\gamma_{a,r}}-\frac{e^{-(\frac{\beta(2^{R_a}-1)-\alpha(2^{R_s}-1)}{\alpha2^{R_s}\bar\gamma_{r,e}})}}{1+\frac{\beta\bar\gamma_{a,r}}
{\alpha\bar\gamma_{r,b}}+\frac{\beta\bar\gamma_{a,r}}{\alpha2^{R_s}\bar\gamma_{r,e}}} \right] \nonumber \\
 +\frac{2^{R_s}\bar\gamma_{r,e}}{2^{R_s}\bar\gamma_{r,e}+\bar\gamma_{r,b}}e^{-(\frac{\beta}{\bar\gamma_{r,b}}+\frac{\beta+1-2^{R_s}}{2^{R_s}\bar\gamma_{r,e}})}
 \left[e^{-\frac{2^{R_a}-1}{\bar\gamma_{a,r}}}-\frac{\frac{\beta\bar\gamma_{a,r}}
{\alpha\bar\gamma_{r,b}}+\frac{\beta\bar\gamma_{a,r}}{\alpha2^{R_s}\bar\gamma_{r,e}}}{1+\frac{\beta\bar\gamma_{a,r}}
{\alpha\bar\gamma_{r,b}}+\frac{\beta\bar\gamma_{a,r}}{\alpha2^{R_s}\bar\gamma_{r,e}}}e^{-\frac{\alpha}{\bar\gamma_{a,r}}}  \right] &\alpha<2^{R_a}-1
\\
\text{the same with (35)} &\alpha\ge 2^{R_a}-1
\end{cases}
\tag{36}
\end{align}
	\hrulefill
	\vspace*{4pt}
\end{figure*}

  and the probability Alice is selected to transmit the message is (32) at the top of the page.
  So the probability that Relay keeps idle is

\begin{align}
\setcounter{equation}{32}
\boldsymbol{\rho}^*_{id}(\! \alpha,\beta \!)\!&\!=\!\begin{cases}
\!1\!\!-\!e^{\!-\!\frac{2^{R_a}-1}{\bar{\gamma}_{a,r}}}\!-\!e^{\!-\!{\frac{\beta}{\bar{\gamma}_{r,b}}}}\!\!+\!e^{\!-\!(\!\frac{\beta}{\bar{\gamma}_{r,b}}\!+\!\frac{2^{R_a}\!-\!1}{\bar{\gamma}_{a,r}}\!)\!}
&\alpha\!<\!2^{R_a}\!-\!1
\\
\text{the same as (24)}&\alpha\!\ge\!2^{R_a}\!-\!1
\end{cases}
\end{align}

\subsection{Performance Analysis For Fixed-rate Transmission Model}
  In this part, the case that Relay may not send real-time CSI feedback and Alice transmits the message with a fixed rate is considered. According to link selection algorithm 1, the probability that Alice is selected to transmit the message is

 \begin{align}
 \setcounter{equation}{30}
\boldsymbol{\rho}^*_a\!(\!\alpha,\beta\!)\! &= \text{Pr}[\gamma_{a,r}[k]\ge max\{\alpha,2^{R_a}-1,\frac{\alpha}{\beta}\gamma_{r,b}[k]\}] \nonumber\\
&=\begin{cases}
\!e^{-\frac{2^{R_a\!}-\!1}{\bar{\gamma}_{a,r}}}\!\!-\!\!\frac{\alpha\bar{\gamma}_{r,b}e^{-(\frac{2^{R_a}-1}{\bar{\gamma}_{a,r}}+\frac{\beta(2^{R_a}-1)}{\alpha\bar{\gamma}_{r,b}})}}{\alpha\bar{\gamma}_{r,b}+\beta\bar{\gamma}_{a,r}} &\alpha<2^{R_a}-1
\\
\text{the same as (22)}&\alpha\ge2^{R_a}-1
\end{cases}
\end{align}
  \par Thus when Alice transmits the data with fixed rate $R_a$, the overall $P_{sop}$ changes to
\begin{align}
\boldsymbol{\rho}_{so}&\triangleq1-(1-{\mathcal{P}^{1*}_{so}})(1-{\mathcal{P}^{2*}_{so}}),
\end{align}
where
\begin{align}
{\mathcal{P}^{1*}_{so}}&=\text{Pr}[R_a\!-\!C_{a,e}\!<\!R_s)|\gamma_{a,r}\ge\text{max}(\alpha,2^{R_a}\!-\!1,\frac{\alpha\gamma_{r,b}}{\beta})]\nonumber \\
&=\text{Pr}[R_a-log_2(1+\gamma_{a,e})<R_s)] \nonumber \\
&=e^{-{\frac{2^{R_a-R_s}-1}{\bar{\gamma}_{a,e}}}},
\end{align}
  and $\mathcal{P}^{2*}_{so}$ is (36) at the top of the page.
  \par Now we reformulate the optimal problem in fixed-rate transmission model and denote them as PF1 and PF2, which are counterparts of the general problem P1 and P2, respectively. Notice that (36)-(41) are related with the relationship between $\alpha, 2^{R_a}-1$, respectively. Hence, the optimal problem is divided into two cases basing on the relationships between $\alpha$ and $2^{R_{a,r}^{data}}-1$, the final result is obtained by comparing the two cases.
  \par Denoting the optimal problems as P-fix1 and P-fix2 in the fixed-rate transmission model with channel outage feedback, which are corresponding to P1 and P2. P-fix1 is formulated as

\begin{equation*}
\begin{split}
\text{PF1:} \mathop{\,\,\text{maximize}}\limits_{\alpha,\beta,R_s} \;\;\;\; &Q(\alpha,\beta,R_s)\\
\text{subject to} \;\;\;\;&\text{C1}:\boldsymbol{\rho}_{so}(\alpha,\beta,R_s) \le \mu \\
                          &\text{C2}:\boldsymbol{\rho}_{id}(\alpha,\beta) \le \nu   \\
                          &\text{C3}:0<R_s<R_{a}\\
                          &\text{C4}:\text{min}(\alpha,\beta)\ge2^{R_s}-1,
\end{split}
\end{equation*}
  Pf2 is

 \begin{equation*}
\begin{split}
 \text{PF2:}\mathop{\,\,\text{minimize}}\limits_{\alpha,\beta,R_s} \;\;\;\; &\boldsymbol{\rho}_{so}(\alpha,\beta,R_s)\\
\text{subject to} \;\;\;\;&\text{C1}:Q(\alpha,\beta,R_s) \ge \theta \\
                          &\text{C2}:\boldsymbol{\rho}_{id}(\alpha,\beta) \le \nu   \\
                          &\text{C3}:0<R_s<R_{a}\\
                          &\text{C4}:\text{min}(\alpha,\beta)\ge2^{R_s}-1
\end{split}
\end{equation*}
  As above introduced, we know that decoding outage event never occur when Alice communicates with Relay. So $\tau_{a,r}$ is

\begin{align}
 \setcounter{equation}{40}
\tau^*_{a,r}&=\mathrm{E}\{\Phi_{a,r}[k]\Gamma_{a,r}[k](1-|I_{k}|)R_s\} \nonumber \\
&=[\gamma_{a,r}\!\ge\!\text{max}(\alpha,2^{R_a}\!\!-\!\!1,\!\frac{\alpha\gamma_{r,b}}{\beta}),R_a\!\!-\!\!log_2(1\!+\!\gamma_{a,e})\!\ge \! R_s)]R_s\nonumber \\
&=\boldsymbol{\rho}^*_a*(1-{\mathcal{P}^{1*}_{so}})R_s.
\end{align}
  and $\tau^*_{r,b}$ (41) at the top of this page.
  So the optimal problem PF3 is

 \begin{equation*}
\begin{split}
 \text{P-fix2:}\mathop{\,\,\text{minimize}}\limits_{\alpha,\beta,R_s} \;\;\;\; &\tau_{r,b}(\alpha,\beta,R_s))\\
\text{subject to} \;\;\;\;&\text{C1}:\boldsymbol{\rho}_{id}(\alpha,\beta) \le \nu   \\
                          &\text{C2}:\tau_{a,r}(\alpha,\beta,R_{s})=\tau_{r,b}(\alpha,\beta,R_s) \\
                          &\text{C3}:0<R_s<R_{a}\\
                          &\text{C4}:\text{min}(\alpha,\beta)\ge2^{R_s}-1
\end{split}
\end{equation*}

\section{NUMERICAL RESULTS AND DISCUSSION} \label{section:simulation}
  \par in this section, first simulation results are given to verify the secrecy outage probability for the proposed link selection schemes and the theoretical derivation of maximum secrecy outage capacity. Then we present numerical examples of the trade-off between SOP and secrecy outage capacity for various values of the constraints.

\begin{figure}[t]
\centering
\includegraphics[width=0.8\linewidth]{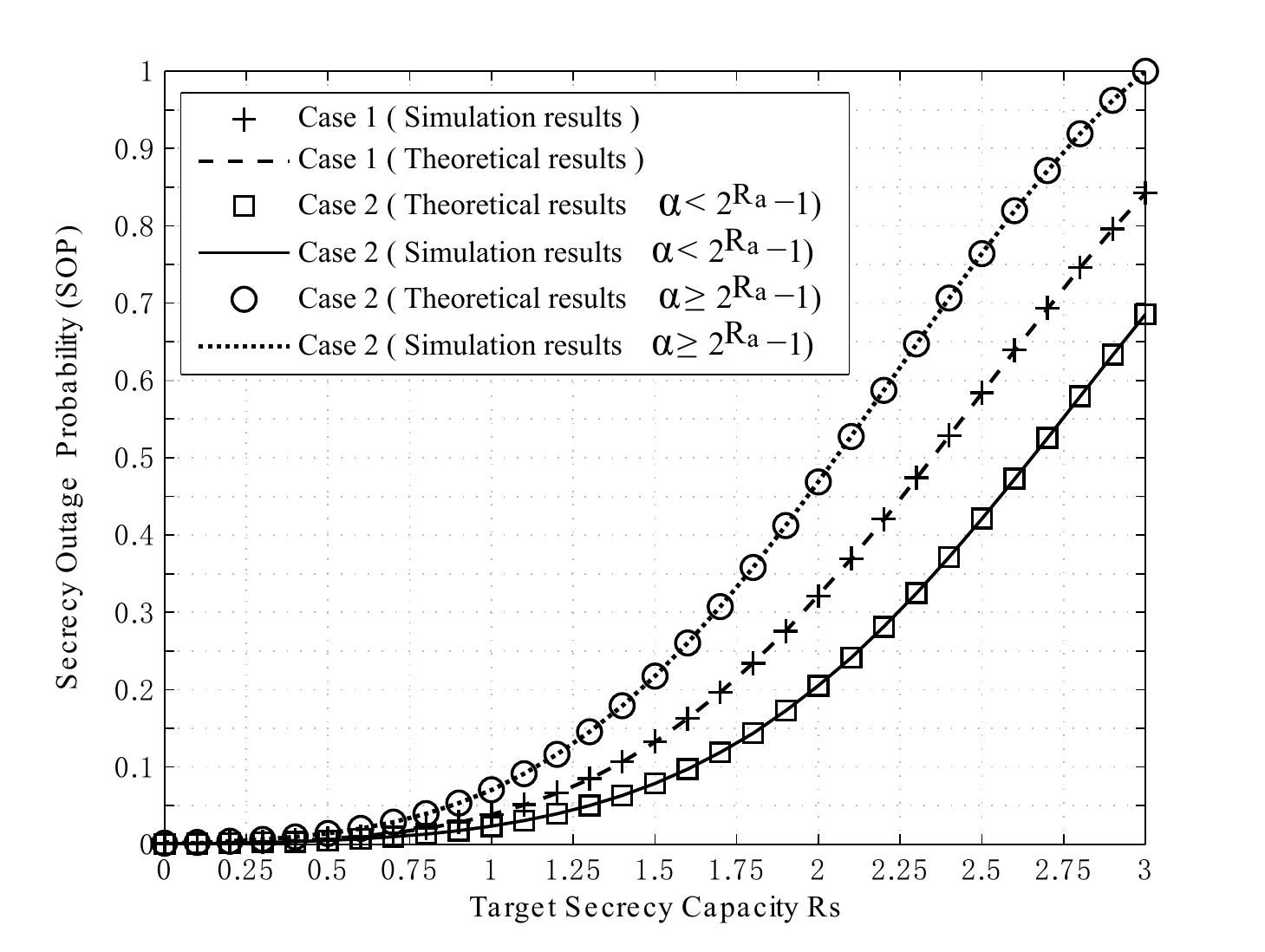}
\caption{Theoretical and simulated End-to-End SOP performance vs. the target secrecy capacity $R_s$. Case 1: adaptive-rate transmission model. Case 2: Fixed-rate Transmission model. }
\label{7}
\end{figure}

\subsection{Simulation Settings and Validation}
  \par For the validation of our theoretical framework, we developed a specific C++ simulator to simulate the system model described in Section \uppercase\expandafter{\romannumeral2} and link selection schemes under adaptive-rate and fixed-rate transmission models.
  Without losing generality, in every below simulations, all noise variance and transmission powers are normalized to unity, and we set the average channel gains of Alice-Relay, Relay-Bob, Alice-Eve and Relay-Eve channels as 5dB, 10dB, 0dB and 2dB, respectively. The duration of each task of simulation is set to be $1 \times 10^{8}$ time slots.
  \par Fig.7 compares the theoretical and simulated values of the End-to-End secrecy outage performance for two network scenarios. In order to void the decoding outage (i.e., $R_s$ is beyond the transmission rate), we respectively set $\alpha=7.0$ and $\beta=8.0$ for adaptive-rate transmission. For fixed-rate transmission, we consider two cases (Case 1: adaptive-rate transmission model. Case 2 fixed-rate transmission model:$\alpha<2^{R_a}-1\,\,\text{and}\,\, R_a=4.0$, case 2:$\alpha\ge2^{R_a}-1\,\, \text{and} \,\,R_a=3.0$). We can see from Fig.7 that the simulation results essentially perfectly match with the theoretical ones. It is clearly shown that case 2 in fixed-rate has the worst outage performance and the case 1 in fixed-rate has the best.

\subsection{Adaptive-rate Transmission Performance Discussion}
\begin{figure}[t]
\centering
\includegraphics[width=0.8\linewidth]{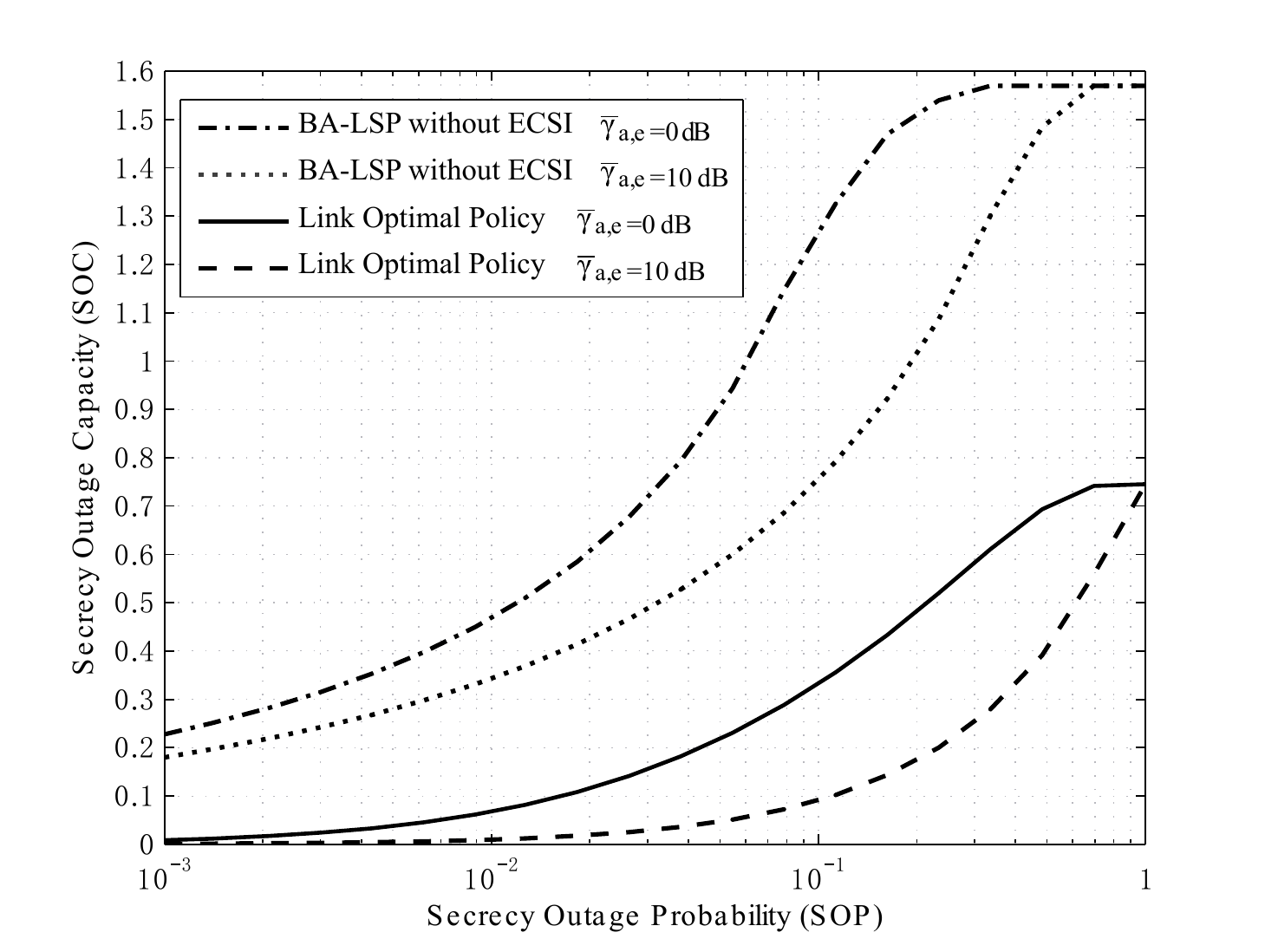}
\caption{Secrecy outage capacity versus End-to-End SOP for $\bar{\gamma}_{a,e}\!\!=\!0,\,10\,\text{dB}$ for different link schemes.}
\label{8}
\end{figure}
  \par In Figure 8, the secrecy outage capacity versus the desired End-to-End SOP constraint for the proposed scheme and link optimal Policy [20] is shown. In order to reveal the effectiveness of the proposed scheme, we consider different eavesdropping channel conditions ($\bar{\gamma}_{a,e}=0, 10\,\,\text{dB}$) in first hop. Figure 8 shows that a higher SOP results in a larger secrecy outage capacity as expected in general for both the proposed scheme and link optimal policy. No matter the equality of the eavesdropping channel in first hop is weak of strong, The proposed link scheme outperforms the link optimal policy. The latter only considers the transmission security in second hop and gives higher priority to select the Alice-relay link with respect to the proposed scheme, so that the buffer queue is at the absorbing state as much as possible.

\begin{figure}[t]
\centering
\includegraphics[width=0.8\linewidth]{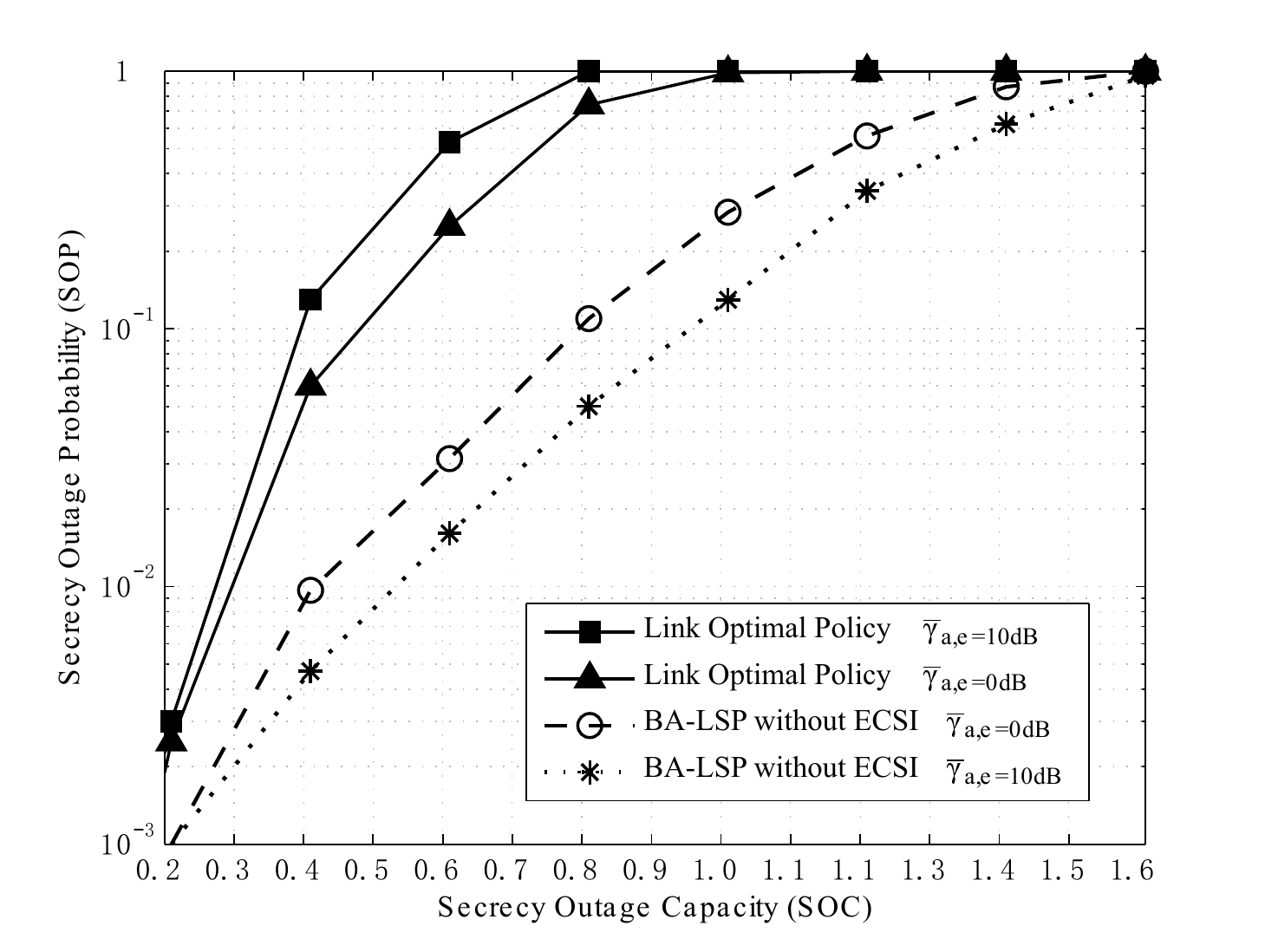}
\caption{End-to-End SOP versus secrecy outage capacity for $\bar{\gamma}_{a,e}\!\!=\!0,\,10\,\text{dB}$ for different link schemes.}
\label{9}
\end{figure}

  \par Figure 9 illustrates the End-to-End SOP versus secrecy outage capacity for different wiretap channel conditions between different schemes. It can be observed that the proposed scheme achieves better secrecy performance under both weak and strong qualities of the wiretap channel. More interesting is that the secrecy outage performance improves more significantly with $\bar{\gamma}_{a,e}=10\,\text{dB}$, it is because that the proposed scheme considers both two wiretap links and the choice of $\alpha,\beta$ has more flexibility. When Alice-Eve link has higher quality, $\alpha$ takes bigger value accordingly and the relay-Bob link is given priority to be selected relatively.

  \par Figure 10 illustrates the exact secrecy throughput versus the probability that the relay keeps idol for different channel gains in second hop. It can be seen that the channel gain in second hop has a positive impact on the secrecy throughput. It is due to the reason that the higher channel gain brings the higher transmission rate and higher rate redundancy, the average number of bits is bigger from the secrecy data in buffer queue of relay to Bob. Thus the exact secrecy throughput from Alice to Bob has bigger value. Another very interesting observation is that relaxing constraint about the idol probability has positive impact on secrecy throughput, which means that Alice and relay are required to transmit the data as much as possible, but the average number of secrecy data that Bob receives is less. This is due to the reason that the effects of secrecy throughput are two folds. On one hand, the small value of idol probability implies higher transmission probabilities for Alice and relay, the probability that the data is intercepted by eavesdropper become higher in dual hops, which leads to the lower secrecy data number both at relay and Bob; on the other hand, a lower idol probability restricts the smaller value ranges of the link selection thresholds $\alpha$ and $\beta$ and the value of the secrecy throughput is related with the value of $\alpha$ and $\beta$.

\begin{figure}[t]
\centering
\includegraphics[width=0.8\linewidth]{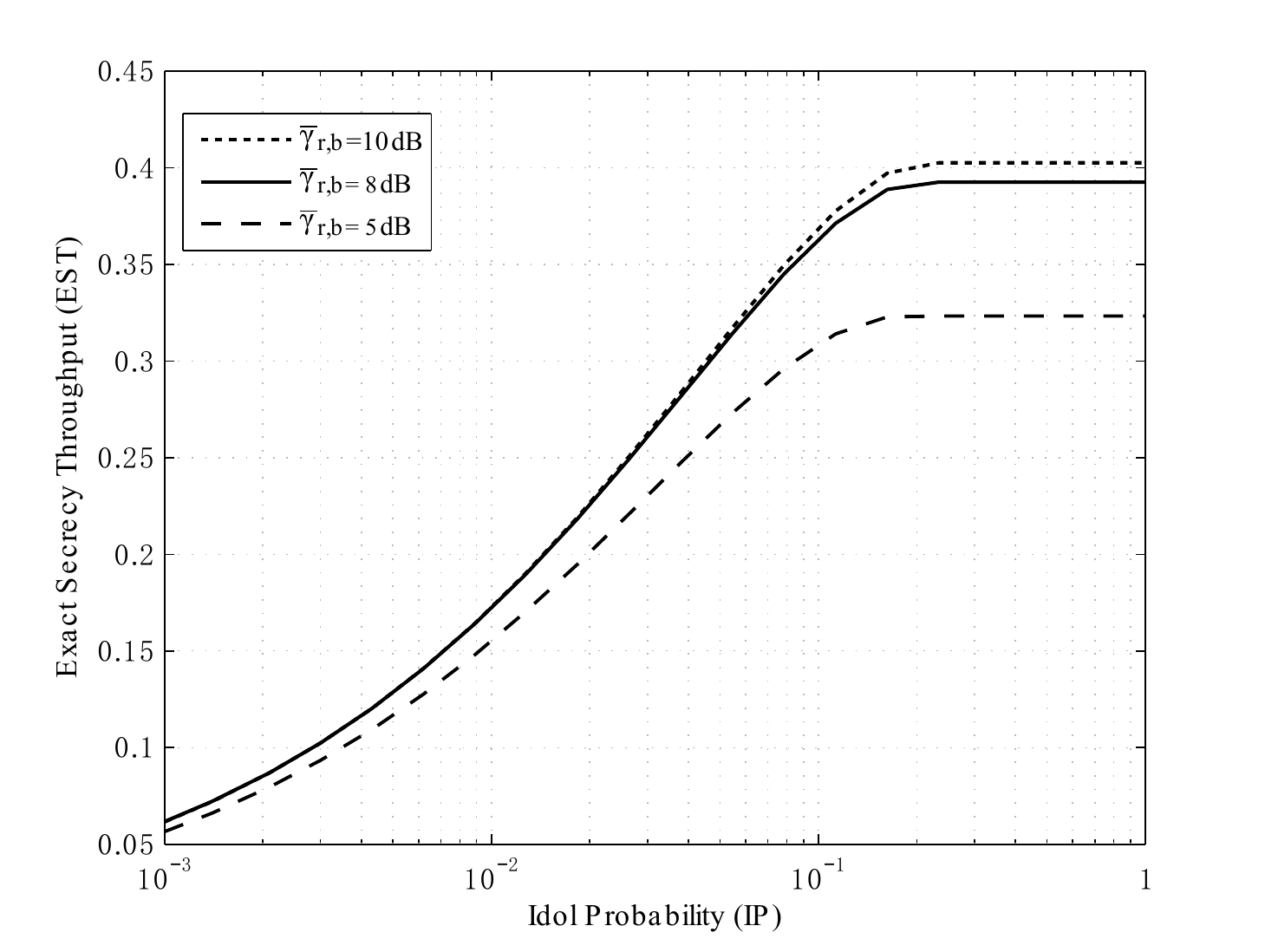}
\caption{Exact secrecy throughput versus idol probability (i.e., the probability that relay neither accepts nor transmits the message consider of two-hop security)}
\label{10}
\end{figure}

\subsection{Fixed-rate Transmission Performance Discussion}

\begin{figure}[t]
\centering
\includegraphics[width=0.8\linewidth]{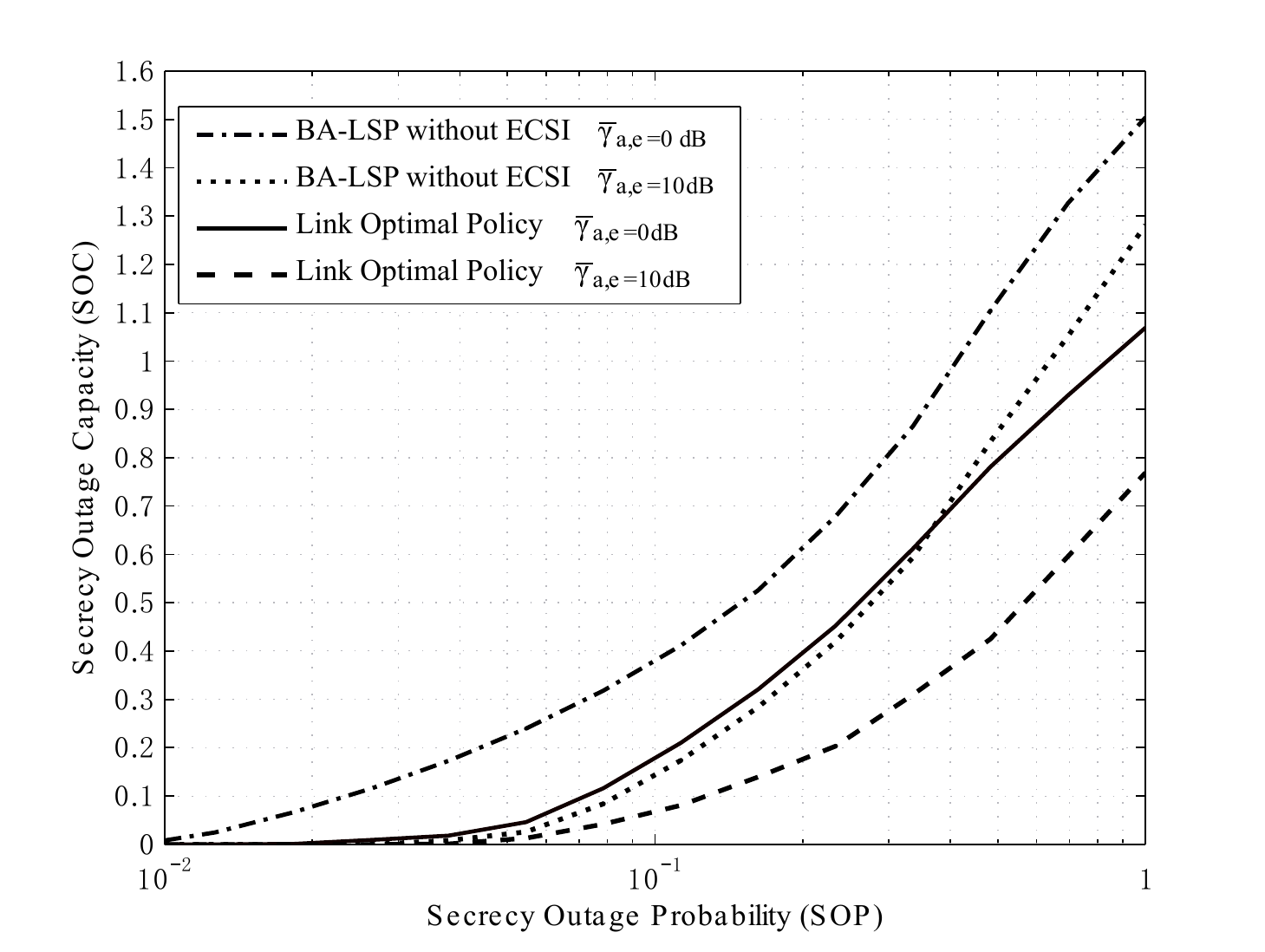}
\caption{Secrecy outage capacity versus End-to-End SOP for $\bar{\gamma}_{a,e}\!\!=\!0,\,10\,\text{dB}$ between different link schemes.}
\label{11}
\end{figure}

  \par Figure 11 depicts the secrecy outage capacity versus the desired End-to-End SOP constraint for the proposed scheme and link optimal Policy in case 2. As expected in general, the secrecy outage capacity increases with the value of SOP growing. Interestingly, for both weak and strong channel quality of the Alice-relay link, we proposed scheme precedes the link optimal policy. Compared with case 1 Figure 8, we see that the secrecy outage capacity performance decreases sharply with the same SOP constraint. Different from case 1, Alice transmits data with fixed-rate $R_a$ in case 2, which leads to the higher SOP in first hop.

\begin{figure}[t]
\centering
\includegraphics[width=0.8\linewidth]{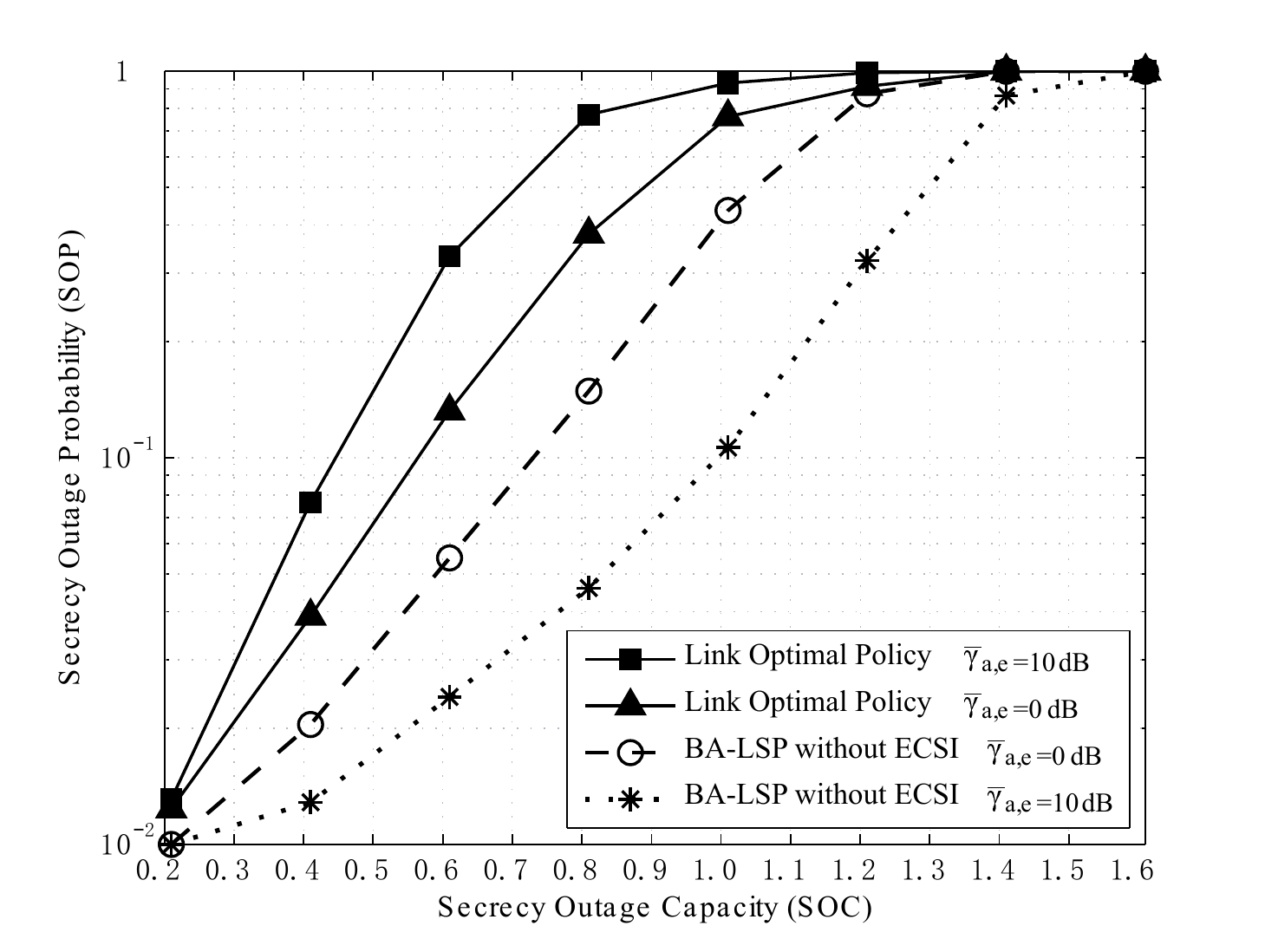}
\caption{The end-to-end SOP versus secrecy outage capacity for $\bar{\gamma}_{a,e}\!\!=\!0,\,10\,\text{dB}$ between different link schemes..}
\label{12}
\end{figure}

  \par In Figure 12, we depict the tradeoff between the End-to-End SOP versus secrecy outage capacity with $\bar{\gamma}_{a,e}\!\!=\!0,\,10\,\text{dB}$ under different link selection schemes. It can be seen that the secrecy performance is worse under more stringent secrecy outage capacity constraint. Similarly, the proposed link selection scheme is superior to the link optimal policy, especially when the eavesdropping channel has better quality in the first hop. Considering the Rayleigh fading effect, the fixed-rate transmission model is always inferior to adaptive-rate transmission model.

\begin{figure}[t]
\centering
\includegraphics[width=0.8\linewidth]{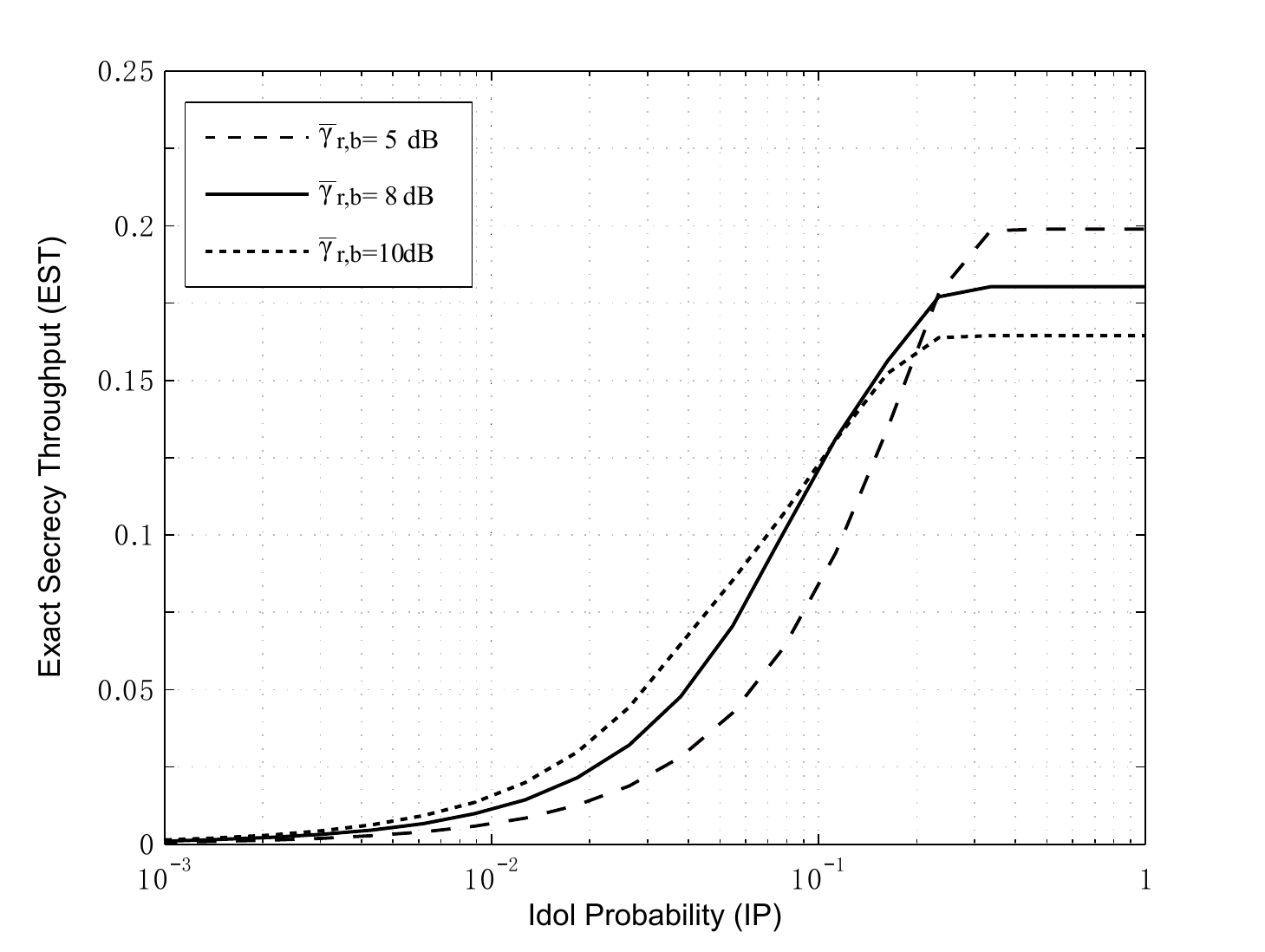}
\caption{Exact secrecy throughput versus idol probability under fixed-rate transmission model.}
\label{13}
\end{figure}

  \par Figure 13 depicts the exact secrecy throughput versus idol probability in case 2 (i.e., fixed-rate transmission), where the final results is obtained by comparing two cases, i.e., $\alpha<2^{R_a}-1$ and $\alpha\ge2^{R_a}-1$. The curves for $\bar{\gamma}_{r,b}=5, 8, 10$ dB show the higher quality of relay-Bob channel results in higher secrecy throughput when $\nu<0.1$, which implies that the delay sensitive network (such as disaster relief and military networks) should give the higher priority when relay-Bob link is higher channel quality but the delay insensitive network  should give lower priority.

\section{Conclusion} \label{section:conclusion}
  This paper considers a two-hop buffer-aided relay network where both the source and relay can be wiretapped by an external eavesdropper. Considering the transmission security of dual hops, we proposed different link selection schemes for different transmission models, which take into account the non-availability of eavesdropper's instantaneous CSI. The link selection decision thresholds are optimal to maximize and minimize the secrecy outage capacity and SOC, respectively. In view of the impact that relay keeps idol on exact secrecy throughput, we study the trade-off between them.Various numerical studies have demonstrated the proposed link scheme provides significant  performance improvement in dual hop eavesdropping network in terms of secrecy outage capacity, secrecy outage probability and exact secrecy throughput.

\appendices
\section{Proof of Proposition 1}
  Denoting the p.d.f.s of $\bar{\gamma}_{a,r}$, $\bar{\gamma}_{r,b}$ and $\bar{\gamma}_{a,e}$ as $f_{\alpha}(\bar{\gamma}_\alpha)$, $f_r(\bar{\gamma}_r)$ and $f_e(\bar{\gamma}_e)$, respectively.

\begin{align}
\tau_{a,r}&=\mathrm{E}\{\Phi_{a,r}[k]\Gamma_{a,r}[k](1-|I_k|)\}R_s \nonumber \\
&=\text{Pr}[log_2\frac{1+\bar{\gamma}_{a,r}}{1+\bar{\gamma}_{a,e}}\ge R_s,\bar{\gamma}_{a,r}\ge\text{max}(\alpha,\frac{\alpha}{\beta}\bar{\gamma}_{r,b})] \nonumber \\
&=\text{Pr}[\bar{\gamma}_{a,r}\ge\text{max}\{\alpha,2^{R_s}(1+\bar{\gamma}_{a,e})-1\},\bar{\gamma}_{r,b}<\beta] \nonumber \\
&\,\,\,\,\,+\text{Pr}[\bar{\gamma}_{a,r}\ge\text{max}\{\frac{\alpha}{\beta}\bar{\gamma}_{r,b},2^{R_s}(1+\bar{\gamma}_{a,e})-1\},\bar{\gamma}_{r,b}\ge\beta] \nonumber \\
&=\int_{0}^{\beta} \left(\int_{0}^{\frac{\alpha+1-2^{R_s}}{2^{R_s}}}\!\!\int_{\alpha}^{\infty}+\int_{\frac{\alpha+1-2^{R_s}}{2^{R_s}}}^{\infty}\int_{2^{R_s}(1+\bar{\gamma}_{a,e})}^{\infty} \right) \nonumber\\
&\,\,\,\,\,f_\alpha(\bar{\gamma}_\alpha)f_e(\bar{\gamma}_e)f_r(\bar{\gamma}_r)d\bar{\gamma}_\alpha d\bar{\gamma}_e d\bar{R_s}{\gamma}_r \nonumber\\
&+\int_{\beta}^{\infty}\!\left(\!\int_{\frac{\alpha}{\beta}\bar{\gamma}_{r,b}}^{\infty} \!\int_{0}^{\frac{\frac{\alpha}{\beta}\bar{\gamma}_{r,b}\!+\!1}{2^{R_s}}-1}\!\!+\!\int_{\frac{\frac{\alpha}{\beta}\bar{\gamma}_{r,b}+1}{2^{R_s}}-1}^{\infty}\int_{2^{R_s}(1+\bar{\gamma}_{a,e})-1}^{\infty}   \!\right)\nonumber\\
&\,\,\,\,\,f_e(\bar{\gamma}_e)f_\alpha(\bar{\gamma}_\alpha)f_r(\bar{\gamma}_r)d\bar{\gamma}_ed\bar{\gamma}_\alpha d\bar{\gamma}_r \nonumber\\
&=u(\alpha,\beta,R_s)+v(\alpha,\beta,R_s),
\end{align}
  where

\begin{align}
u(\alpha,\beta,R_s)&=e^{-\frac{\alpha}{\bar{\gamma}_{a,r}}}(1-e^{-\frac{\beta}{\bar{\gamma}_{r,b}}})\left[1-\frac{2^{R_s}\bar{\gamma}_{a,e} e^{\frac{2^{R_s}-\alpha-1}{\bar{\gamma}_{a,e}2^{R_s}}}}
{\bar{\gamma}_{a,r}+2^{R_s}\bar{\gamma}_{a,e}} \right],
\end{align}
  and

\begin{align}
v(\alpha,\beta,R_s)&=\frac{\beta\bar{\gamma}_{a,r}}{\beta\bar{\gamma}_{a,r}+\alpha\bar{\gamma}_{r,b}}e^{-(\frac{\beta}{\bar{\gamma}_{r,b}}+\frac{\alpha}{\bar{\gamma}_{a,r}})}\nonumber\\
&=\frac{\bar{\gamma}_{a,e}2^{R_s}}{\bar{\gamma}_{a,e}2^{R_s}+\bar{\gamma}_{a,r}}\frac{e^{-(\frac{\alpha}{\bar{\gamma}_{a,r}}+\frac{\beta}{\bar{\gamma}_{r,b}}+
\frac{\alpha+1-2^{R_s}}{\bar{\gamma}_{a,e}2^{R_s}})}}{1+\frac{\alpha\bar{\gamma}_{r,b}}{\beta\bar{\gamma}_{a,r}}+\frac{\alpha\bar{\gamma}_{r,b}}{\beta2^{R_s}\bar{\gamma}_{a,e}}}.
\end{align}

\bibliographystyle{IEEEtran}
\bibliography{IEEEabrv,reference}

\end{document}